\documentclass[twocolumn]{aastex62}
\usepackage{graphicx,ifthen,url,float,color,ulem,amsmath,ragged2e}

\newcommand {\etal}{et.~al.}
\newcommand {\kms}{km s$^{-1}$}

\def\ltsima{$\; \buildrel < \over \sim \;$}
\def\simlt{\lower.5ex\hbox{\ltsima}}
\def\gtsima{$\; \buildrel > \over \sim \;$}
\def\simgt{\lower.5ex\hbox{\gtsima}}
\newcommand {\uJy}{$\mu$Jy}
\newcommand {\um}{$\mu$m}

\newcommand{\sfr}{{\rm\,M$_\odot$\,yr$^{-1}$}}
\newcommand{\lsun}{{\rm\,L$_\odot$}}
\newcommand{\lstar}{{\rm\,L$_\star$}}
\newcommand{\phistar}{{\rm\,$\Phi_\star$}}

\newcommand{\lir}{L$_{\rm IR}$}
\shorttitle{ALMA Deep Fields}
\shortauthors{Casey et al.}

\begin{document}

\title{\sc An Analysis of ALMA Deep Fields and the Perceived Dearth of High-$z$ Galaxies}

\correspondingauthor{Caitlin M. Casey}
\email{cmcasey@utexas.edu}

\author[0000-0002-0930-6466]{Caitlin M. Casey}
\affil{Department of Astronomy, The University of Texas at Austin, 2515 Speedway Blvd Stop C1400, Austin, TX 78712}

\author{Jacqueline Hodge}
\affil{Leiden Observatory, Niels Bohrweg 2, 2333 CA Leiden, The Netherlands}

\author[0000-0002-7051-1100]{Jorge A. Zavala}
\affil{Department of Astronomy, The University of Texas at Austin, 2515 Speedway Blvd Stop C1400, Austin, TX 78712}

\author[0000-0003-3256-5615]{Justin Spilker}
\affil{Department of Astronomy, The University of Texas at Austin, 2515 Speedway Blvd Stop C1400, Austin, TX 78712}

\author[0000-0001-9759-4797]{Elisabete da Cunha}
\affil{Research School of Astronomy and Astrophysics, The Australian National University, Canberra ACT 2611, Australia}

\author[0000-0002-8437-0433]{Johannes Staguhn}
\affiliation{NASA Goddard Space Flight Center, Code 665, Greenbelt, MD 20771}
\affiliation{Bloomberg Center for Physics and Astronomy, Johns Hopkins University 3400 N. Charles Street, Baltimore, MD 21218}

\author[0000-0001-8519-1130]{Steven L. Finkelstein}
\affil{Department of Astronomy, The University of Texas at Austin, 2515 Speedway Blvd Stop C1400, Austin, TX 78712}

\author[0000-0003-3627-7485]{Patrick Drew}
\affil{Department of Astronomy, The University of Texas at Austin, 2515 Speedway Blvd Stop C1400, Austin, TX 78712}


%

%



\begin{abstract}
Deep, pencil-beam surveys from ALMA at 1.1--1.3\,mm have uncovered an
apparent absence of high-redshift dusty galaxies, with existing
redshift distributions peaking around $z\sim1.5-2.5$.  This has led to
a perceived dearth of dusty systems at $z\simgt4$, and the conclusion,
according to some models, that the early Universe was relatively
dust-poor.  In this paper, we extend the backward evolution galaxy
model described by Casey \etal\ (2018) to the ALMA regime (in depth
and area) and determine that the measured number counts and redshift
distributions from ALMA deep field surveys are fully consistent with
constraints of the infrared luminosity function (IRLF) at $z<2.5$
determined by single-dish submillimeter and millimeter surveys
conducted on much larger angular scales ($\sim$1--10\,deg$^2$).  We
find that measured 1.1--1.3\,mm number counts are most constraining
for the measurement of the faint-end slope of the IRLF at $z\simlt2.5$
instead of the prevalence of dusty galaxies at $z\simgt4$.  Recent
studies have suggested that UV-selected galaxies at $z>4$ may be
particularly dust-poor, but we find their millimeter-wave emission
cannot rule out consistency with the Calzetti dust attenuation law,
even by assuming relatively typical, cold-dust ($T_{\rm
  dust}\approx30$\,K) SEDs.  Our models suggest that the design of
ALMA deep fields requires substantial revision to constrain the
prevalence of $z>4$ early Universe obscured starbursts. The most
promising avenue for detection and characterization of such early
dusty galaxies will come from future ALMA 2\,mm blank field surveys
covering a few hundred arcmin$^2$ and the combination of existing and
future dual-purpose 3\,mm datasets.
\end{abstract}

\keywords{galaxies: starburst -- ISM: dust -- cosmology: dark ages --
  surveys}


\section{Introduction} \label{sec:intro}

Since its commissioning in 2011, the Atacama Large Millimeter Array
(ALMA) has swung open new discovery space in almost every area of
astrophysics.  Its unparalleled sensitivity to tracers of gas and
dust emission, both in the nearby and distant Universe, have
been revolutionary: from intricate gaps in protoplanetary disks around
young stars \citep[e.g.][]{alma-partnership15a,andrews16a}, 
ubiquitous gas outflows from
dense cores of nearby galaxies
\citep{leroy15a,meier15a,ando17a},
 the regular detection of molecular gas and dust in normal massive
 galaxies out to high-redshift \citep{hodge13a,hodge16a,brisbin17a},
dark matter substructure around
massive high-$z$ galaxies \citep{hezaveh13a,hezaveh16a,hezaveh16b}
to the discoveries of the highest-redshift
dusty-star forming galaxies (DSFGs) to-date
\citep{vieira13a,strandet17a,marrone17a}.

One of the key goals of extragalactic work with ALMA has been the
blind survey of the early Universe in dust and gas, to reveal the
nature of obscured emission from an unbiased point of view,
without the guidance of tracers selected at other wavelengths,
primarily the rest-frame ultraviolet or optical.  Dust emission can be
traced directly in submm/mm continuum, while gas can be traced either
indirectly through dust continuum
\citep{scoville14a,scoville16a,scoville17a} or directly through
molecular line transitions like CO
\citep{neri03a,tacconi06a,tacconi08a,casey11b,bothwell13c}, which
allows a three-dimensional mapping of the Universe with both spatial
and spectral data \citep{decarli14a,decarli16a,decarli16b}. 

 It was never quite clear what would be found with blank-field surveys
 by ALMA given how few measurements had ever been made previously
 \citep[and most of those had been done with single-dish submm
   facilities with much larger beamsizes, obfuscating multiwavelength
   counterpart identification;][]{smail97a,barger98a,hughes98a}.  The
 potential for groundbreaking discovery was nevertheless high, given
 our disparate knowledge of the population of galaxies well-studied in
 the optical and near-infrared, and those discovered at submm/mm
 wavelengths.  The two populations often exhibit completely orthogonal
 physical characteristics, from their star-formation rates
 \citep{chapman05a,wardlow11a,gruppioni13a} to their obscuration
 fractions \citep{pannella09a,pannella15a,whitaker14a,whitaker17a},
 while also exhibiting some troubling degeneracies, like optical
 color, which can cause one population to seem indistinguishable from
 another
 \citep{goldader02a,burgarella05a,buat05a,howell10a,takeuchi10a,casey14a}.

Thus, the first several years of ALMA operation has seen the initial
results of the first ALMA blind pencil-beam surveys, including both
blank dust-continuum detection experiments
\citep{dunlop16a,hatsukade16a,aravena16a,franco18a}, molecular gas
deep fields \citep{walter16a,decarli16a,decarli16b}, and blank
dust-continuum follow-up around specially-chosen protocluster fields
\citep{umehata15a}.  One common result among these surveys has been
the relative dearth of faint sources discovered at high-redshift
($z>4$).  This paper address why that might be the case, focusing
exclusively on galaxies' dust-continuum emission.  We also synthesize
results of prior single-dish work and lessons learned about the
infrared galaxy luminosity function (the `IRLF') to inform future ALMA
deep field campaigns.  This paper draws on a complex backward
evolution model
built to understand and interpret the submm sky, summarized in
\citet{casey18a}, hereafter C18.  This paper specifically explores the
application of this model to ALMA observations.  In \S~\ref{sec:model}
we briefly summarize the model setup, \S~\ref{sec:existing} presents
the results of the model in comparison with existing ALMA deep fields,
\S~\ref{sec:bouwens} presents an alternate analysis of the dust
properties of rest-frame UV-selected galaxy populations, and
\S~\ref{sec:future} comments on the potential discriminating power of
future ALMA deep surveys for refining constraints on the high-$z$
IRLF.  We assume a {\it Planck} cosmology throughout this paper,
adopting $H_{0}=67.7\,$\kms\,Mpc$^{-1}$ and $\Omega_{\lambda}=0.6911$
\citep{planck-collaboration16b}.

\section{Model Parameterization}\label{sec:model}

We have built a backward-evolution model to interpret the origins of
emission in the submillimeter/millimeter sky from galaxy number
counts, redshift distributions and correlations between bands.  This
model is built to constrain the nature of the IRLF out to
high-redshifts, where only small handfuls of dust-obscured sources
have been directly characterized. Existing datasets can,
nevertheless, inform our interpretation of those epochs through
statistical comparisons.  A more detailed description of the model's
motivation and structure are provided in C18.  We provide only a brief
summary here.

\begin{figure}
\includegraphics[width=0.99\columnwidth]{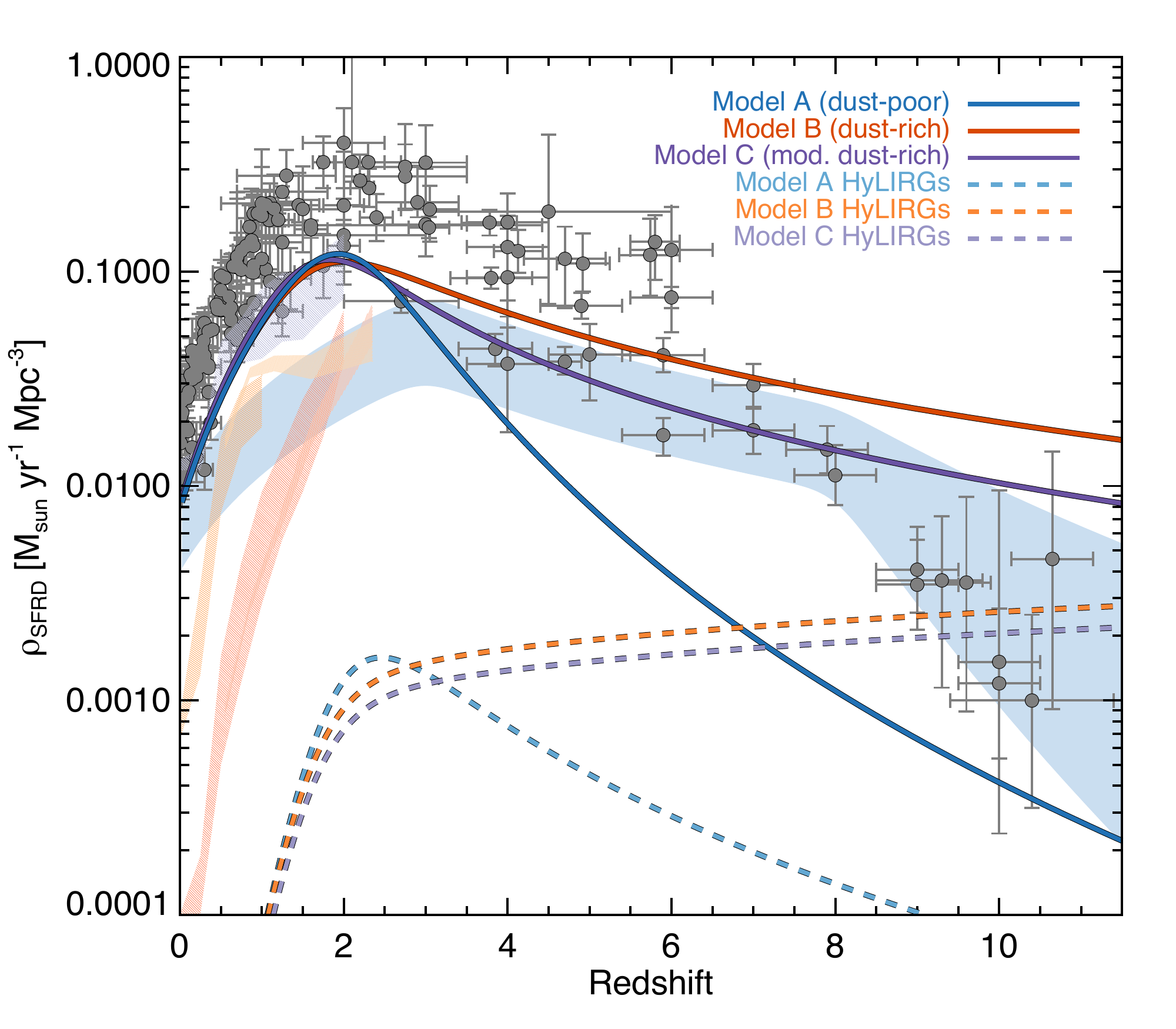}
\caption{The cosmic star-formation rate density as measured across
  multiple literature datasets, as summarized in \citet[][gray
    points]{madau14a}.  The light blue shaded region highlights the
  total measured contribution of unobscured light (rest-frame UV and
  optical tracers).  Light purple, orange and red transparent regions
  represent the measured constraints on total obscured contribution,
  contribution from LIRGs ($10^{11}<L_{\rm IR}<10^{12}$\,\lsun), and
  ULIRGs ($10^{12}<L_{\rm IR}<10^{13}$\,\lsun), respectively.  Thick
  solid lines illustrate the total contribution from obscured galaxies
  as proposed by our three models: Model A, the dust-poor early
  Universe (blue), Model B, the dust-rich early Universe (orange), and
  Model C, the modified dust-rich early Universe (purple).  The
  contribution of HyLIRGs ($L_{\rm IR}>10^{13}$\,\lsun) to each model
  is shown in dashed lines to illustrate that Model B and Model C are
  very similar at the bright-end of the luminosity function.}
\label{fig:sfrd}
\end{figure}

The model first constructs an infrared galaxy luminosity function,
$\Phi(L,z)$, spanning $0<z<12$ with IR luminosities from
$10^{8}<L<10^{14}$\,\lsun\footnote{In the model we abbreviate $L_{\rm
    IR(8-1000\mu\!m)}$ as $L$.}.  At low redshifts this is informed by
direct measurements of the IRLF
\citep{sanders03a,le-floch05a,casey12b,gruppioni13a,magnelli13a}.  At
$z\simgt2.1$, we adopt two possible models for the evolution of the
luminosity function where it is no longer constrained by data.  Both
assume \lstar$\propto(1+z)$.  Model A assumes a very low number
density of DSFGs in the early Universe such that
\phistar$\propto(1+z)^{-5.9}$, following the fall-off in bright
UV-luminous galaxies at the same epoch, while Model B assumes a much
shallower relation, \phistar$\propto(1+z)^{-2.5}$, implying a much
higher prevalence of DSFGs in the early Universe.  
  Figure~\ref{fig:sfrd}, as well as Figure~6 of C18, highlights the
differences in the cosmic star-formation rate densities implied by
either model; model A (the dust-poor Universe) implies that obscured
galaxies might only contribute $\sim$10\%\ toward cosmic
star-formation at $z\simgt 4$ while model B (the dust-rich Universe)
implies that obscured galaxies dominate with $\sim$90\%\ of cosmic
star-formation at $z\simgt 4$.  

The adopted models of the IRLF in C18 fix both the bright-end slope
($\beta_{\rm LF}=-3$) and faint-end slope ($\alpha_{\rm LF}=-0.6$) of
the double powerlaw across all epochs.  This is done due to our lack
of ability to break degeneracies between evolving faint-end slopes and
different evolutions in \lstar\ or \phistar. Also, single-dish surveys
are largely unable to detect galaxies below \lstar\ at high-$z$, and
therefore the prescriptions for the faint-end slope are largely
irrelevant\footnote{Though sub-\lstar\ galaxies can be directly
  detected in the low-redshift Universe, their number density is
  significantly lower than their high-$z$ cousins, and so they
  contribute very little to number counts or redshift distributions.}.

\begin{figure}
\centering
\includegraphics[width=0.99\columnwidth]{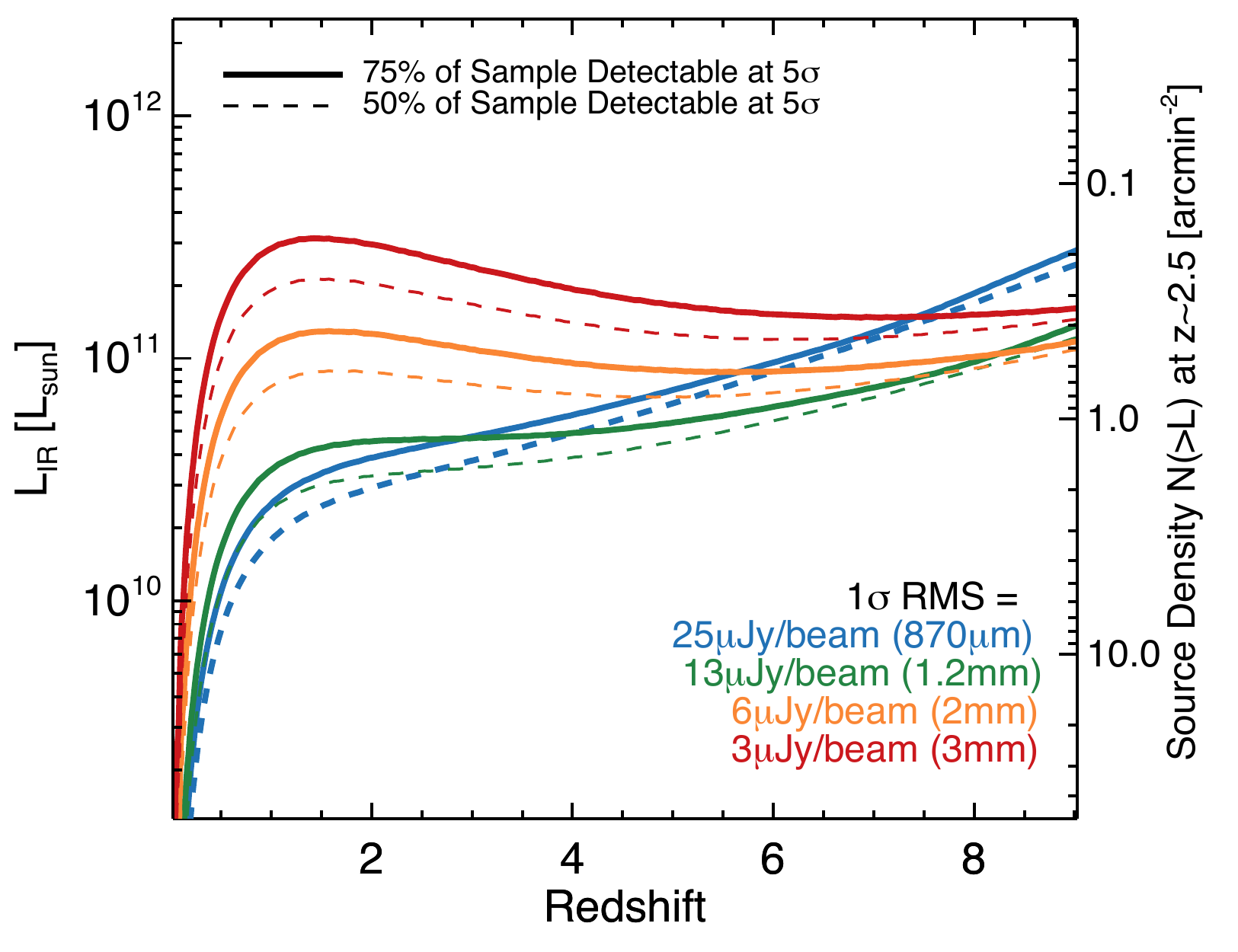}
\caption{ The luminosity sensitivity limits of four
    different ALMA deep field surveys, with 1$\sigma$ RMS depths of
    25\uJy/beam (at 870\um, blue), 13\uJy/beam (at 1.2\,mm, green),
    6\uJy/beam (at 2\,mm, orange) and 3\uJy/beam (at 3\,mm, red) as
    outlined in Table~\ref{tab:setup}.  Line type denotes what
    fraction of the population at the given luminosity and redshift
    would likely be detectable above the given threshold:
    $>$75\%\ (solid) or $>$50\%\ (dashed).  The curves are determined
    by the observed $L_{\rm IR}-\lambda_{\rm peak}$ relationship (see
    Figure~3 of C18) with typical 10\%\ scatter.  At high-redshifts we
    incorporate the impact of CMB heating \citep{da-cunha13a} on
    luminosity detection limits, which effectively flattens out the
    dramatic negative K-correction seen in the millimeter beyond
    $z\sim6$.  The right y-axis is labeled with the approximate source
    density of sources above a given luminosity on the sky at
    $z\approx2.5$.  We discuss the important trade-offs of survey area
    vs. depth later in the paper.}
\label{fig:lirz}
\end{figure}

Beyond the adopted functional form of the luminosity function, our
model then assigns an infrared spectral energy distribution
(3\um--3\,mm) to individual sources according to a probability density
function that is dependent on the source's integrated IR luminosity
$L$ and redshift $z$.  SED rest-frame peak wavelengths are a function
of $L$ at each redshift, such that more luminous galaxies are
intrinsically hotter. We find no significant evidence for an evolution
in the $L$-$\lambda_{\rm peak}$ relationship.  See Figure~3 of C18 and
the discussion in \S~2.2 for details on how the SEDs are generated.
Our SEDs are parameterized via $\lambda_{\rm peak}$ instead of dust
temperature $T$ of the ISM, which makes the model insensitive to
different opacity assumptions that impact the relationship between the
observable $\lambda_{\rm peak}$ and the physical quantity\footnote{As
  will be shown in Figure~\ref{fig:seddiff}, optically thin
  vs. optically thick assumptions can dramatically impact a galaxy's
  SED with fixed dust temperature.  This motivates our focus on
  rest-frame peak wavelength, $\lambda_{\rm peak}$, instead of dust
  temperature itself.  While the observables might change
  substantially at a fixed temperature, they do not for a fixed
  $\lambda_{\rm peak}$.}  $T$.  However, the impact of heating from
the cosmic microwave background (CMB) at very high redshifts is a
strong function of the underlying physical dust temperature $T$
\citep{da-cunha13a}.  As in C18, and informed by the rough luminosity
sensitivities of ALMA deep field surveys shown in
Figure~\ref{fig:lirz}, we continue with the assumption that SEDs
transition from optically thick to thin with $\tau=1$ at 100\um.

With luminosity functions and SEDs in-hand, sources are then injected
into mock maps at any wavelength along the SED.  A mock
  map consists of a regularly-spaced grid with pixel size equal to 1/5
  of the minimum beamsize FWHM simulated; positions of injected
  sources are randomly assigned.  In C18, we investigated the
characteristics of maps spanning the IR through millimeter, from
70\um\ through 2\,mm.  Once all sources at all redshifts have been
injected into these mock maps, they are convolved with the beamsize of
observations specific to a certain instrument at a certain
observatory, instrumental noise is added to the maps, and sources are
re-extracted to compare against real observations.

\begin{figure*}
\centering
\includegraphics[width=1.9\columnwidth]{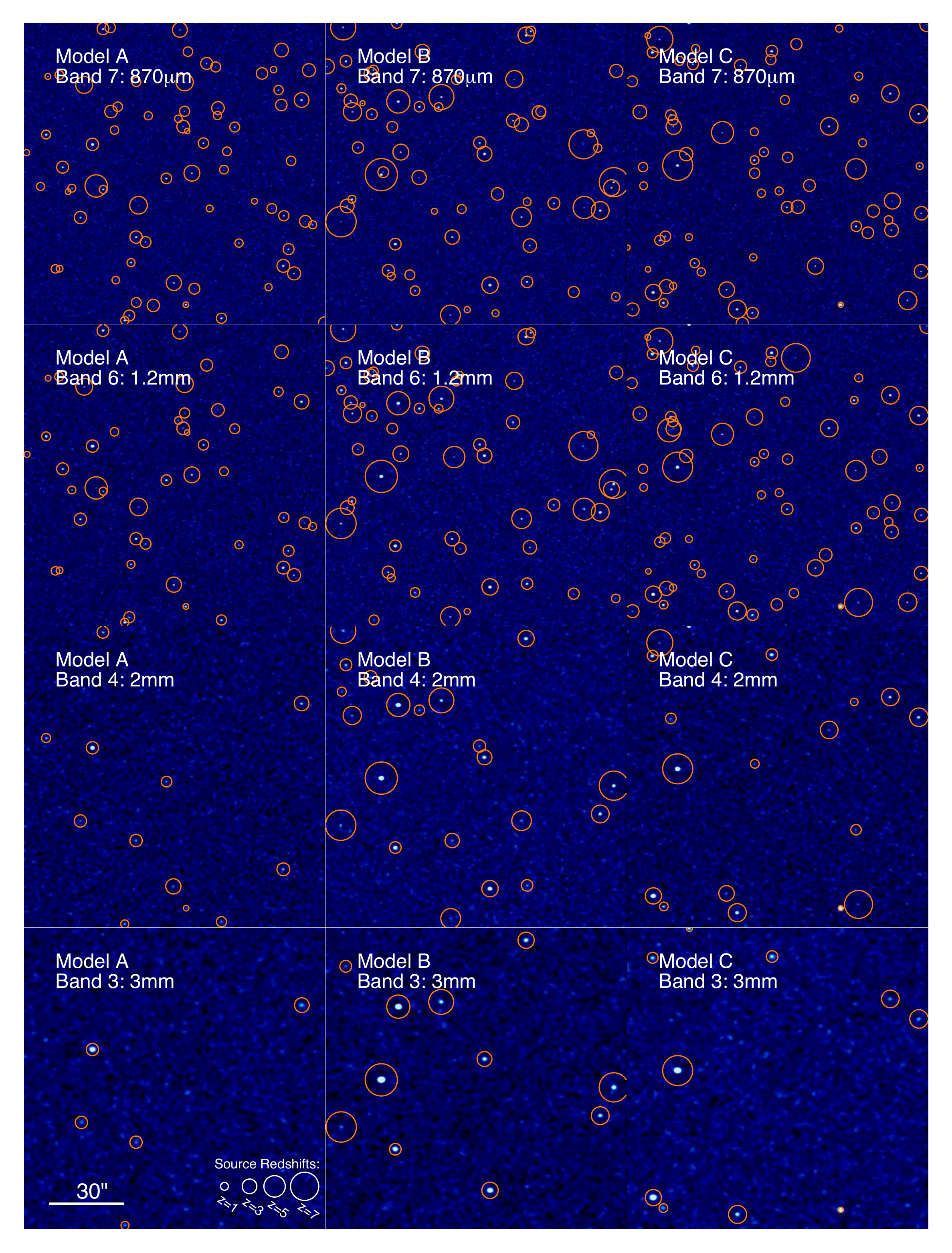}
\caption{2$'\times$2$'$ cutouts of mock ALMA maps at 870\um\ (Band 7;
  top row), 1.2\,mm (Band 6; second row), 2\,mm (Band 4; third row)
  and 3\,mm (Band 3; last row).  The left column represents the output
  from Model A, the dust-poor Universe model.  The middle column is
  the output from Model B, and the right column from Model C; both
  Models B and C represent a dust-rich Universe model, with different
  prescriptions for the faint-end slope of the luminosity function.
  The assumed RMS noise values for these maps are given in
  Table~\ref{tab:setup}. Sources detected at $>$5$\sigma$ significance
  are encircled in orange in all maps; for illustrative purposes, the
  circle size is proportional to injected source redshift (a legend is
  given in the low left panel).  The full redshift distributions for
  all samples is given in Figure~\ref{fig:zdist}.}
\label{fig:cutouts}
\end{figure*}

In this paper, we draw up mock ALMA deep field maps in band 6 using
the quoted beamsizes and RMS noise values of ALMA campaigns---though
small adjustments to the beamsize are negligible since our maps are
not confusion-limited.  Since we do not model the galaxies' sizes
directly, and instead input them as point sources, adjustments to the
angular resolution on the order of 0.5$''$--2$''$ do not change our
results, but we do note that sources extended on $>$0.5$''$ are
somewhat common \citep{hodge16a} and the potential to resolve sources
at higher angular resolution should be taken into account for
designing future observational campaigns.  In addition to the modeled
1.2\,mm maps (band 6), we simulate hypothetical maps at 870\um\ (band
7), 2\,mm (band 4) and 3\,mm (band 3) to interpret what role they
might play in constraining dust emission at high-$z$.
Table~\ref{tab:setup} lists the observational setups we test in this
paper, and Figure~\ref{fig:lirz} shows the rough
  luminosity limits of these flux density thresholds in the four ALMA
  bands.  Figure~\ref{fig:cutouts} shows mock maps at all sample
wavelengths given each setup.

\begin{table}
\caption{Characteristics of Observational Setups}
\centering
\begin{tabular}{cccc}
\hline\hline
{\sc Passband} & {\sc Instrument}/ & {\sc Beamsize} & {\sc RMS} \\
               & {\sc Telescope}   &  FWHM [$''$]   & [\uJy]     \\
\hline
870\um\ & {\sc ALMA Band 7} & 0.5$\times$0.5 & 25 \\
1.2\,mm & {\sc ALMA Band 6} & 0.6$\times$0.6 & 13 \\
2\,mm   & {\sc ALMA Band 4} & 1.0$\times$1.0 & 6 \\
3\,mm   & {\sc ALMA Band 3} & 1.5$\times$1.5 & 3  \\
\hline\hline
\end{tabular}
\label{tab:setup}

{\small {\bf Notes.} This table summarizes the different observational
  setups we test for on 1--400\,arcmin$^2$ scales in our ALMA-focused
  simulations.  The chosen beamsizes and RMS values are typical of
  observations available in the ALMA archive at each frequency.}
\vspace{2mm}
\end{table}

\subsection{Importance of the Faint-End Slope of the IRLF}

It is clear from a number of quick tests on the C18 model that it is
$\alpha_{\rm LF}$, the faint-end slope of the luminosity function,
that has the most profound and dominating effect on the density of
sources in 1.2\,mm ALMA deep fields.  Because this paper focuses on
these deep fields, which probe a bit deeper than the single-dish
results summarized in C18, we expand on the C18 models A and B in this
paper by also testing different values for $\alpha_{\rm LF}$.

\begin{figure}
\includegraphics[width=0.99\columnwidth]{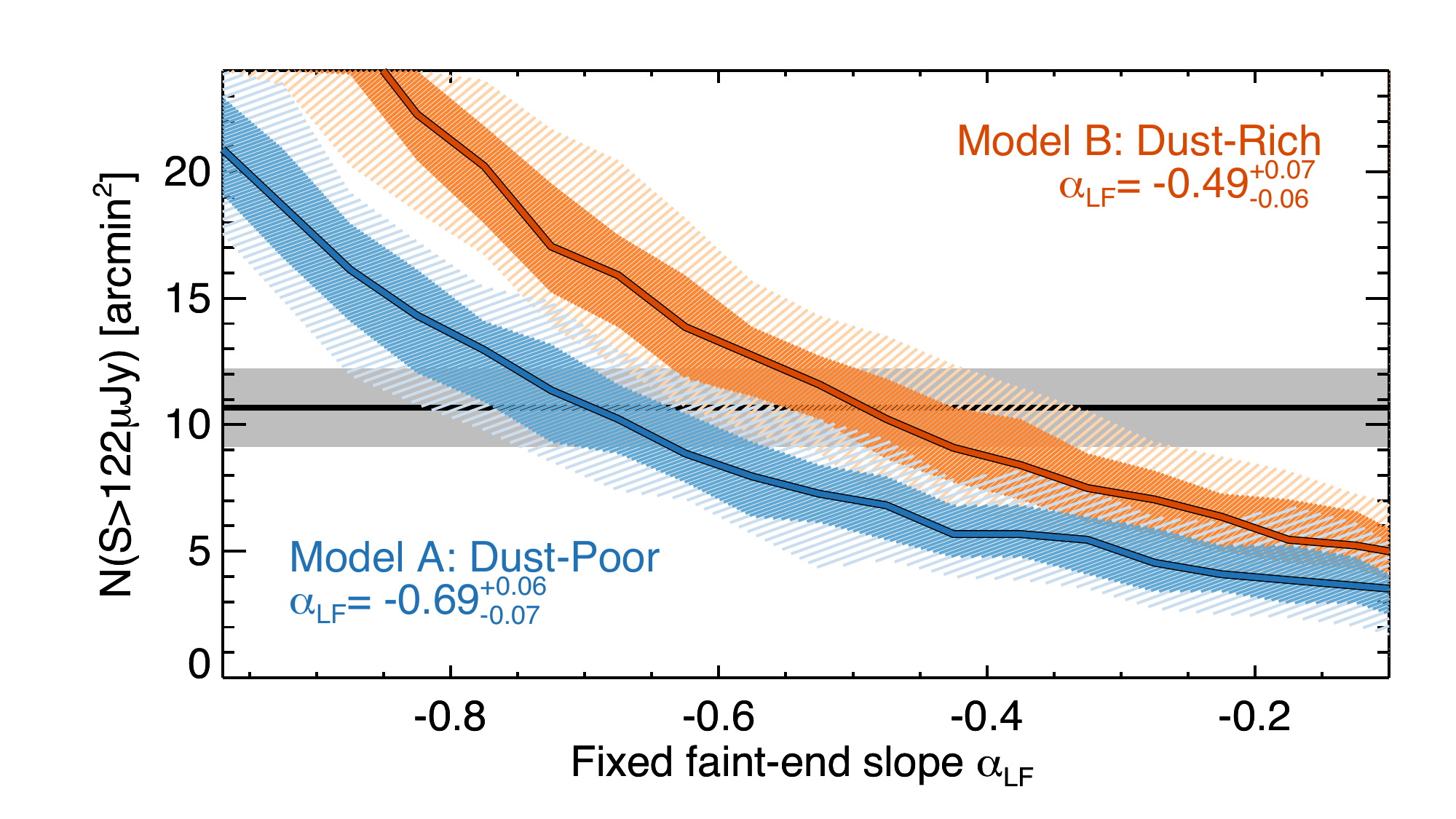}
\caption{The number of sources per arcmin$^2$ found in our simulated
  maps as a function of the faint-end slope of the luminosity
  function, $\alpha_{\rm LF}$.  The black horizontal line and gray
  error region denotes the measured number of sources per square
  arcminute in the \citet{dunlop16a} 4.4\,arcmin$^2$ HUDF map; 47
  sources were identified above a 3.5$\sigma$ significance with a
  1$\sigma$ RMS of 35\uJy.  At a fixed value of $\alpha_{\rm LF}$ we
  simulate 100 such 4.4\,arcmin$^2$ maps and identify the number of
  $>$3.5$\sigma$ sources.  The blue line and error region show the
  results of Model A, the dust-poor early Universe, while the orange
  line denotes the results of Model B, the dust-rich early Universe.
  C18 assumes a fixed faint-end slope of $\alpha_{\rm LF}=-0.6$, while
  here we measure best agreement with the HUDF dataset if $\alpha_{\rm
    LF}=-0.69^{+0.06}_{-0.07}$ for Model A and $\alpha_{\rm
    LF}=-0.49^{+0.07}_{-0.06}$ for Model B.}
\label{fig:alphafix}
\end{figure}
In Figure~\ref{fig:alphafix}, we show the results of adjusting the
value of the faint-end slope (within $-1<\alpha_{\rm LF}<-0.1$) for
both Models A and B.  All other parameters in the models are fixed to
the values as given in Table~3 of C18.  This figure shows the number
of detected sources above 3.5$\sigma$ significance with a
35\,\uJy\ RMS as a function of $\alpha_{\rm LF}$ \citep[these values
  follow the specifications of ][where 47 sources are identified above
  this threshold in a 4.4\,arcmin$^2$ map]{dunlop16a}.  At a fixed
value of $\alpha_{\rm LF}$, we constrain the number of expected
sources and its uncertainty by using 100 Monte Carlo simulations for
either Model A or Model B.  At fixed $\alpha_{\rm LF}$, Model A will
produce 30\%\ fewer sources than Model B, directly attributable to the
different adopted values of $\psi_{2}$, the parameter determining the
high-$z$ evolution of \phistar.  This indicates that $\sim$70\%\ of
sources in our simulated maps are likely to sit at redshifts
unaffected by model differences, mainly $z<2$.  We explore this more
fully in the next section.  The values of $\alpha_{\rm LF}$ that
result in best agreement with the HUDF source density are $\alpha_{\rm
  LF}=-0.69^{+0.06}_{-0.07}$ for Model A and $\alpha_{\rm
  LF}=-0.49^{+0.07}_{-0.06}$ for Model B.

In what follows, we analyze a few different permutations of the models
as a result of the impact of the faint-end slope of the luminosity
function.  For illustrative purposes, we continue our analysis of
Model A and Model B exactly as given in C18, fixing the faint-end
slope to $\alpha_{\rm LF}=-0.6$.  In addition, we also provide
analysis of Model A with its best-fit value of $\alpha_{\rm LF}=-0.69$
and Model B with its best-fit value of $\alpha_{\rm LF}=-0.49$.  We
also introduce a Model C in this paper, which is a modification of the
dust-rich Model B.  The only change from Model B is that Model C
allows the faint-end slope to evolve with redshift like:
\begin{equation}
\alpha_{\rm LF}= \left\{
\begin{array}{lr}
  \alpha_{\rm 0}(1+z)^{a_{1}} & : z\ll z_{\rm turn}\\
  \alpha_{\rm 0}(1+z)^{a_{2}} & : z\gg z_{\rm turn}\\
\end{array}
\right.
\label{eq:alphaevol1}
\end{equation}
Physically, this model is motivated by the observed steepening of the
rest-frame UV-slope towards the highest redshifts
\citep{bouwens07a,bouwens15a,reddy09a,mclure13a,finkelstein15a} and
also in the steepening of the low-mass end of the stellar mass
function \citep{grazian15a,duncan15a,song16a}.  The IRLF might
logically exhibit the opposite behavior by flattening at increased
redshift.  In other words, this promotes the idea that low-mass
galaxies should be less dust-enhanced at earlier redshifts than at
later redshifts.
Following the methods of C18, the adopted functional form is then
dependent on $x\equiv\log_{10}(1+z)$, $x_{t}=\log_{10}(1+z_{\rm
  turn})$ and $x_{w}=z_{w}/(1+z_{\rm turn})$, where $z_{w}\equiv2.0$,
such that:
\begin{equation}
\begin{split}
\log\alpha_{\rm LF}(x) = & -\frac{(a_2-a_1)x_{w}}{2\pi} \Big[\ln\big(\cosh(\pi\frac{x-x_{t}}{x_{w}})\big)\\
                     & - \ln\big(\cosh(-\pi\frac{x_{t}}{x_{w}})\big)\Big]\\
                     & - \frac{(a_2-a_1)}{2}x - \log(-\alpha_{\rm 0})\\
\end{split}
\label{eq:alphaevol}
\end{equation}
For Model C we adopt all of the same parameters as Model B of C18
(e.g. $z_{\rm turn}=1.8$, $z_{\rm w}=2.0$).  While Model B would have
both $a_{1}$ and $a_{2}$ set to zero in Equations~\ref{eq:alphaevol1}
and \ref{eq:alphaevol} we set $a_{1}=0$ and $a_{2}=-0.7$ to
accommodate a flattening of the faint-end of the luminosity function
at high-redshift.  We set $\alpha_{\rm 0}=-0.69$ at $z=0$, in line
with the measured best-fit value for Model A from
Figure~\ref{fig:alphafix}.

In summary, we analyze the results of three different model universes
in this paper.  The first is Model A, the dust-poor Universe model,
that assumes very few DSFGs beyond $z>4$, while the second is Model B,
the dust-rich Universe model, that assumes DSFGs make up
$\sim$90\%\ of the cosmic star-forming budget at $z>4$. Both models A
and B explore different values of $\alpha_{\rm LF}$, either fixed to
$-$0.6 as in C18 or, for most analysis in this paper, adjusted to the
best-fit data-driven value as found in Figure~\ref{fig:alphafix}.
The variation of $\alpha_{\rm LF}$ values motivates the introduction
of Model C.  Model C is a modified version of Model B; mainly, it
proposes a dust-rich early Universe with high prevalence of DSFGs at
high-redshifts, but fewer and fewer lower luminosity DSFGs with
increasing redshifts resulting in a flatter slope to the IRLF.  Note
that all three models provide plausible fits to the number counts of
galaxies at higher flux densities as measured from single-dish surveys
(see C18).

\begin{figure*}
\centering
\includegraphics[width=0.99\columnwidth]{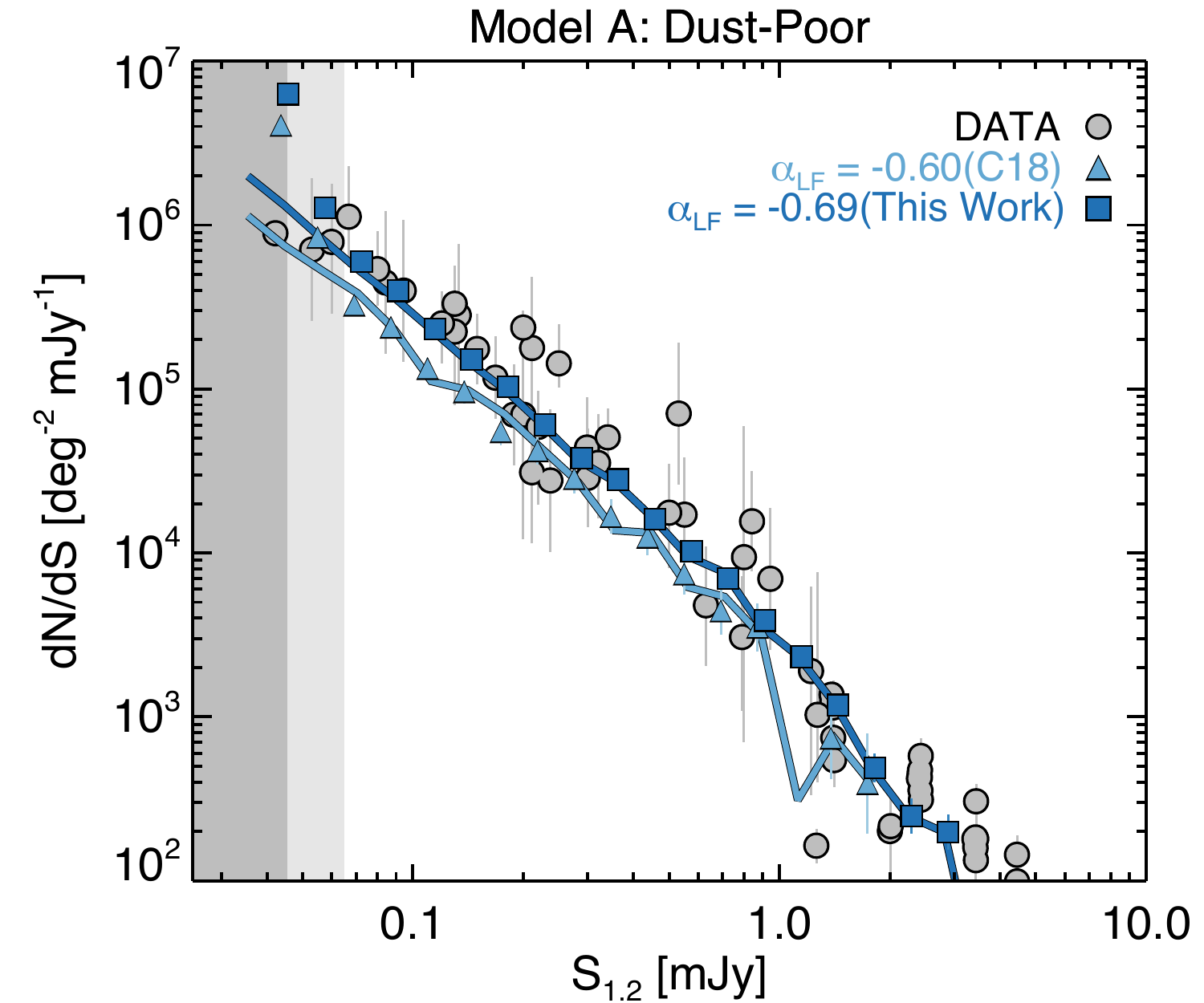}\\
\includegraphics[width=0.99\columnwidth]{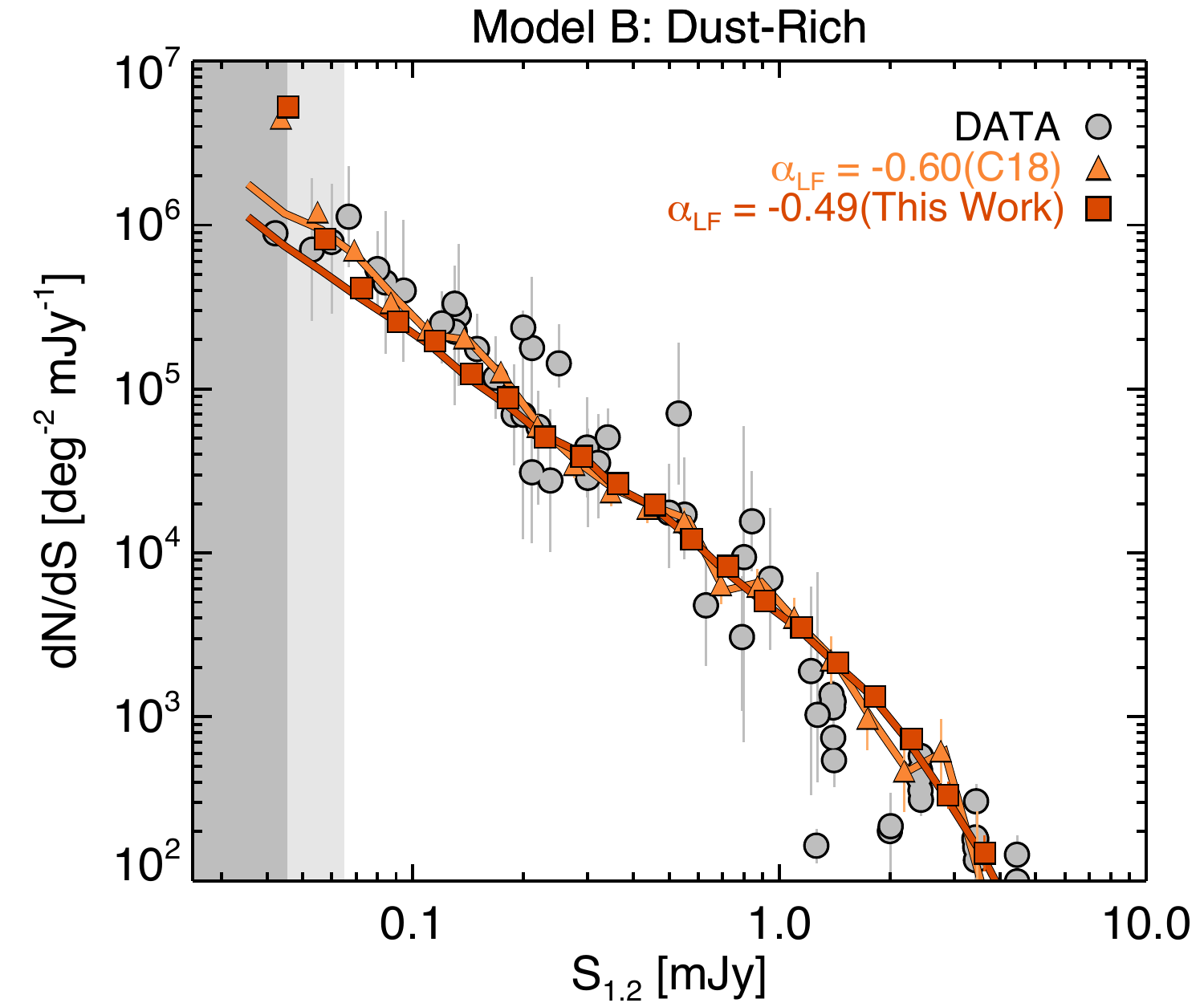}
\includegraphics[width=0.99\columnwidth]{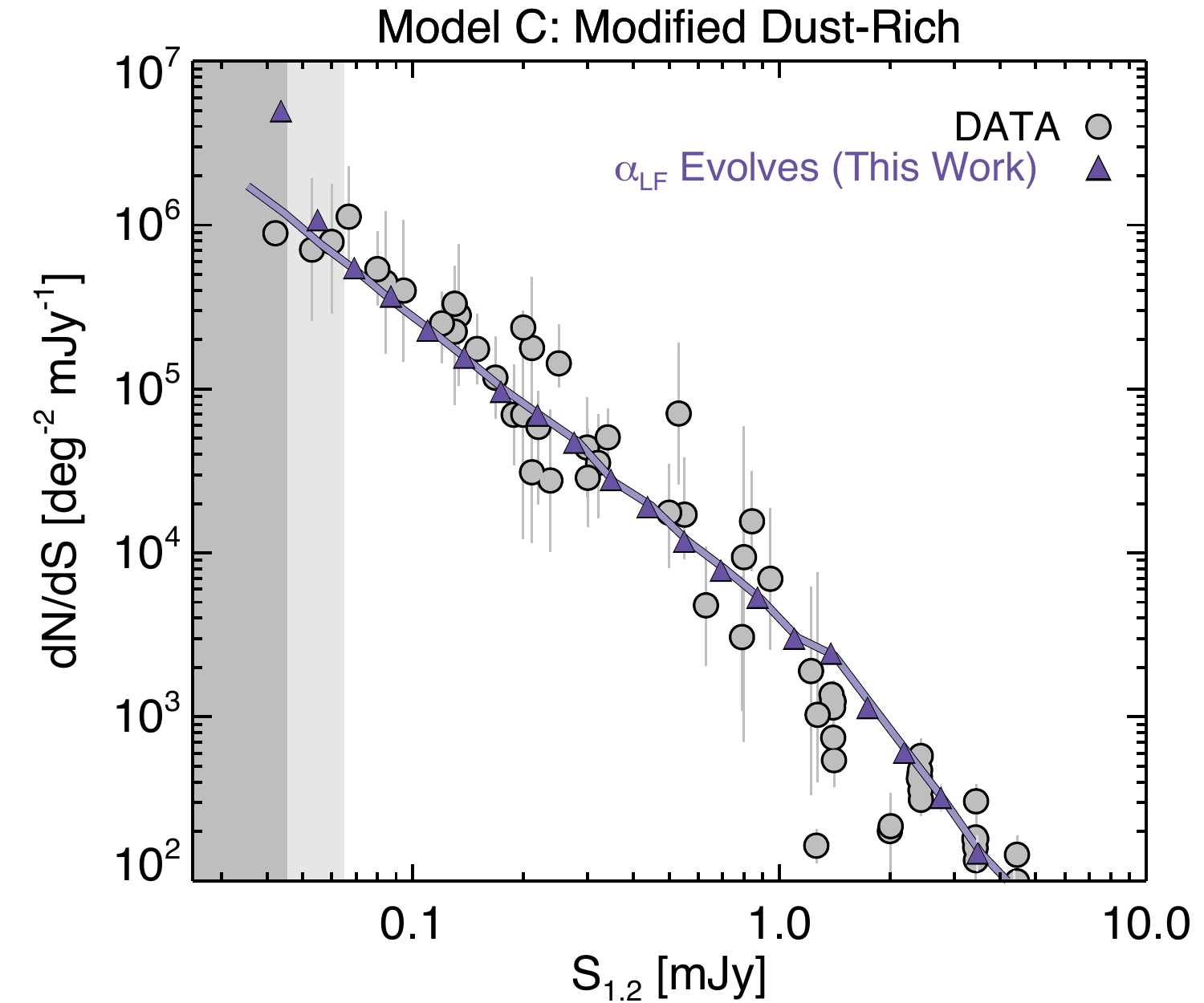}
\caption{A comparison of 1.2\,mm number counts from the literature
  \citep[gray
    points;][]{ono14a,carniani15a,dunlop16a,hatsukade16a,aravena16a,fujimoto16a,oteo16a,franco18a},
  and our simulations output.  Sources are extracted down to
  3.5$\sigma$ significance; the dark gray region represents flux
  densities at $<$3.5$\sigma$, while the light gray region represents
  sources with $3.5<\sigma<5$.  The contamination rate below 5$\sigma$
  exceeds 10\%, and so we advocate for analysis of individual systems
  only above 5$\sigma$.  At top, we show the family of models that
  assume a dust-poor early Universe, with
  \phistar$\propto(1+z)^{-5.9}$ (Model A). Below, we assume a
  dust-rich early Universe with \phistar$\propto(1+z)^{-2.5}$ (Model B
  and Model C).  Both Model A and Model B assume a fixed faint-end
  slope of the luminosity function ($\alpha_{\rm LF}=-0.6$) as in C18
  (triangles), and then re-measure the number counts using the
  best-fit faint-end slope as measured in Figure~\ref{fig:alphafix}
  (squares).  Model C is a variant of Model B where $\alpha_{\rm LF}$
  is allowed to evolve such that the IRLF slope at the faint end
  becomes shallower with increasing redshift.  All injected source
  counts are shown as solid lines, while extracted source counts shown
  as symbols (triangles or squares).  This figure shows that all
  models (A--C) agree with measured number counts at 1.2\,mm --
  despite quite significant differences in assumed number density of
  high-$z$ dusty galaxies.}
\label{fig:nc}
\end{figure*}

\section{Comparison to Existing 1.2\,mm Datasets}\label{sec:existing}

ALMA deep fields have extended our knowledge of millimeter number counts
into the sub-mJy regime, not probed by prior datasets.
These ALMA deep field efforts include:
\begin{itemize}
\vspace{-2mm}
\item SSA22 Core Deep Field \citep{umehata15a}: a 4.5\,arcmin$^2$
  1.1\,mm survey of the $z\sim3.09$ protocluster core to a depth of
  70\,\uJy/beam with a 0.53$\times$0.50$''$ beam,
\vspace{-2mm}
\item The Hubble Ultra Deep Field \citep[HUDF;][]{dunlop16a}: a
  4.5\,arcmin$^2$ 1.3\,mm survey with 0.7$''$ beam to a depth of
  35\,\uJy/beam,
\vspace{-2mm}
\item The SXDF ALMA Deep field \citep{hatsukade16a}: a 2.0\,arcmin$^2$
  survey at 1.1\,mm to a depth of $\sim$55\,\uJy/beam,
\vspace{-2mm}
\item The ASPECS Pilot Deep field \citep{walter16a,aravena16b}: a
  0.79\,arcmin$^2$ 1.2\,mm deep field to a continuum depth of
  12.7\,\uJy/beam and a 1.5$\times$1.0$''$ beam.  ASPECS also mapped
  the same region in 3\,mm continuum, which achieved a 1$\sigma$ RMS
  of 3.8\,\uJy/beam with a 2$\times$3$''$ beam, and
\vspace{-2mm}
\item The GOODS-ALMA Survey \citep{franco18a}: a 69\,arcmin$^2$
  1.1\,mm deep field centered on CANDELS, containing the HUDF pointing
  of \citeauthor{dunlop16a} mapped to an RMS of 0.18\,mJy/beam
  analyzed with a synthesized beam of 0.6$''$ but originally mapped at
  high spatial resolution with a beamsize of 0.2--0.3$''$.
\end{itemize}
\vspace{-1mm} Further observational efforts are currently underway,
primarily the cycle 4 ASPECS large program intended to cover an area
of 4.6\,arcmin$^2$ to a depth similar to the ASPECS pilot survey.
This would significantly deepen the HUDF pointing; note that the
ASPECS-Pilot, HUDF, and GOODS-ALMA surveys are all sequentially nested
in the same patch of sky centered on the HUDF.  Several additional
projects have combined results from archival ALMA datasets to infer
deep (sub)millimeter number counts, including \citet{ono14a} who
measure number counts at 1.1--1.3\,mm down to 0.1\,mJy covering an
area of 3\,arcmin$^2$ across 10 different fields, \citet{carniani15a}
who measure number counts down to 0.1\,mJy across 18 different fields
from the archive ($\sim$4\,arcmin$^2$), \citet{fujimoto16a} who
combined data spanning 10\,arcmin$^2$ of various depths at 1.1\,mm,
and \citet{oteo16a}, which describes the ALMACAL project, which uses
data from the regions around commonly-used ALMA calibrators to produce
deep maps at 870\um\ ($\sim$6\,arcmin$^2$) and 1.1\,mm
($\sim$16\,arcmin$^2$).  Because the vast majority of data collected
is from ALMA band 6 (1.1--1.3\,mm), we restrict our comparative
analysis to the band 6 datasets.

Figure~\ref{fig:nc} shows a detailed comparison of measured number
counts at 1.2\,mm from these literature resources against our three
models.  The solid lines indicate the injected source number counts,
while the colored symbols are the output extracted number counts.
There is remarkable global agreement of the measured number counts
from ALMA and the model output, even in the case of the fixed
$\alpha_{\rm LF}=-0.6$ models from C18, despite the fact that these
models were generated to fit luminosity functions of much brighter
sources found in single-dish surveys only.  Unfortunately the measured
number counts are highly uncertain and susceptible to cosmic variance
due to small number statistics in small, pencil-beam surveys.  It is
not even immediately obvious that the adjustments made to the
faint-end slope of the luminosity function have made a discernible
difference with such substantial scatter from the data themselves.

\subsection{Completeness, Contamination, and Sample Cleaning}

Note that the final sample sizes of the \citeauthor{dunlop16a},
\citeauthor{aravena16a} and \citeauthor{franco18a} works were 16, 9,
and 20 respectively.  Our models predict anywhere between 31--47,
21--38, or 85--146 for the given areas, depths, and detection
thresholds, respectively.  There is some tension in these estimates,
despite no disagreement between their calculated 1.1--1.3\,mm number
counts and our model results (shown in Figure~\ref{fig:nc}).  Here we
explore reasons for this tension in their final source lists.

We caution that disagreement between our model predictions (31--47)
and the \citet{dunlop16a} statistics (16 sources) are likely caused by
the additional cuts that \citeauthor{dunlop16a} make to reduce their
original sample of 47 detections above a 3.5$\sigma$ threshold to 16
with OIR counterparts.  These cuts are motivated by the estimated
contamination rates from spurious sources at the 3.5$\sigma$ detection
threshold.  The raw number of $>$3.5$\sigma$ sources found in
\citeauthor{dunlop16a} is in agreement with our predictions.

Our predictions for the ASPECS-Pilot 0.79\,arcmin$^2$ map (21--38
sources predicted) are discrepant with observations (9 sources
observed).  This tension could be quickly alleviated by adopting a
shallower faint-end slope to the IRLF or by invoking cosmic variance
on such small areas ($<$1\,arcmin$^2$).  Even small variations in the
faint-end slope $\alpha_{\rm LF}$ can have profound effects on the
predicted source counts in such a small, deep drill survey.  For
example, adopting $\alpha_{\rm LF}=-0.6$ (instead of $-0.69$) for
Model A results in a predicted number of sources 40\%\ lower in a mock
ASPECS-Pilot map.  If we modify the detection threshold to 5$\sigma$,
up from 3.5$\sigma$, we note that our predictions, of detecting 5--12
sources, fall more in line with the observed 5 sources.  Indeed, we
find the signal to noise threshold to be rather impactful on the
estimated source contamination rates.

We estimate contamination in our simulations in two ways.  First, we
invert the maps and run our source detection algorithm on the negative
image, where we know all detections will be false. This is an often
used analysis technique for real data, where simulations like ours are
not immediately available.  It should come as no surprise (and is
indeed reassuring) that our detection rate for inverted sources is
uniform across all of our models; we estimate a false rate of
4.6$\pm$0.6 sources per arcmin$^2$.  \citeauthor{dunlop16a} find 29
such spurious sources in their 4.4\,arcmin$^2$ map, which is in
3$\sigma$ tension with our findings, though they also offer other
calculations which estimate $\approx$20 false sources.  Twenty false
sources above $>$3.5$\sigma$ would be in perfect alignment with our
model output.  In the case of the ASPECS-Pilot project, we estimate
3.6$\pm$0.5 false sources in their 0.79\,arcmin$^2$ (out of the 9
sources above 3.5$\sigma$). If we consider the two of their sources
without OIR counterparts as possible contaminants, this agrees nicely
within Poisson uncertainty. However, it does not preclude other false
identifications in the \citet{aravena16a} sample.

\begin{figure}
\includegraphics[width=0.99\columnwidth]{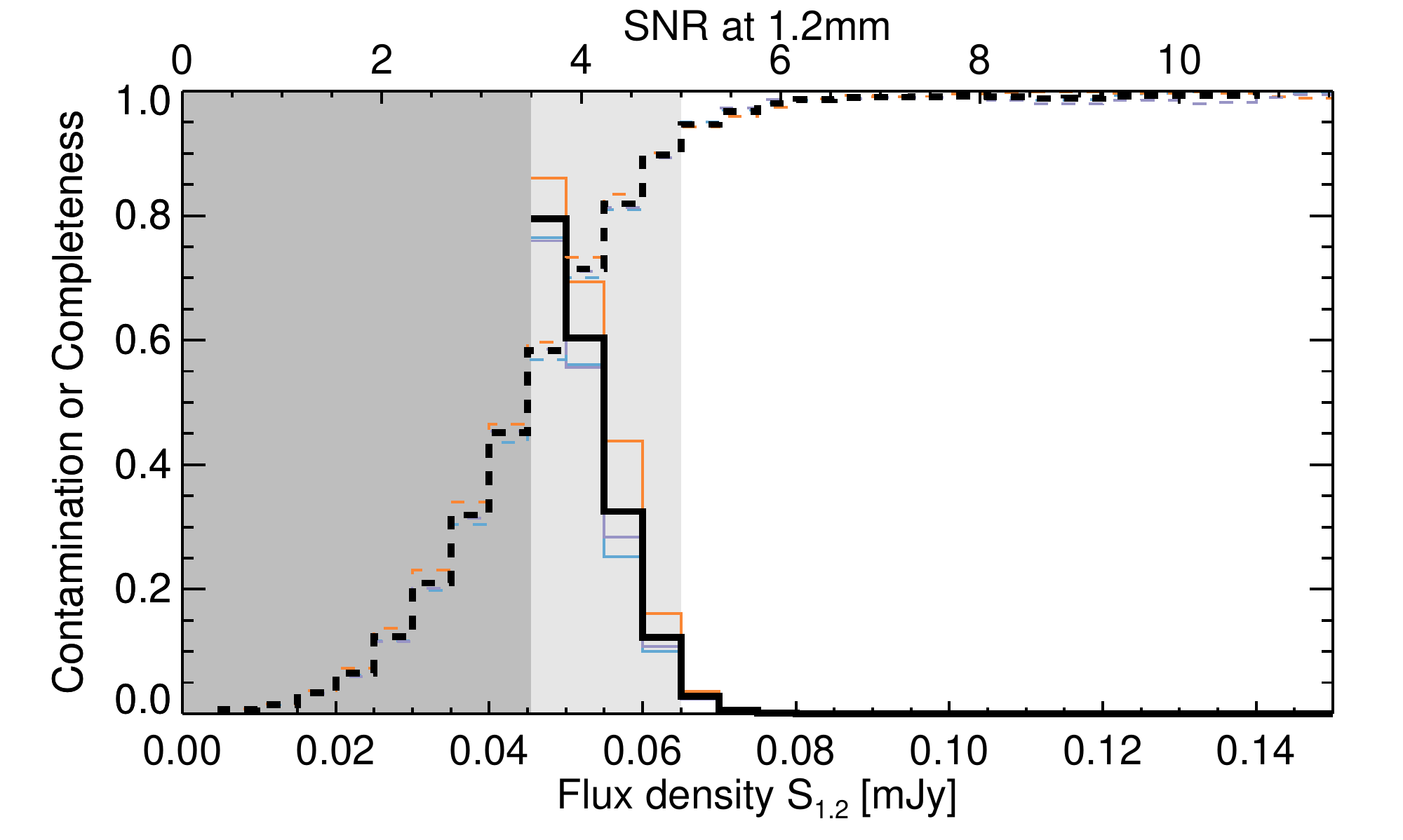}
\caption{The measured contamination (solid lines) and completeness
  (dashed lines) for our simulated 1.2\,mm ALMA maps.  The colored
  lines indicate individual simulations with different model
  assumptions, following the same color scheme as other plots in this
  manuscript.  We do not find any variation by model parameters
  because these maps are not confusion limited.  The average
  contamination and completeness for all simulations is shown in
  black.  We find that sources with $3.5\sigma<$SNR$<5\sigma$ are
  potentially very highly contaminated by positive noise fluctuations,
  and similarly, suffer from $>$10\%\ incompleteness.}
\label{fig:cc}
\end{figure}

\begin{figure*}
\centering
\includegraphics[width=1.2\columnwidth]{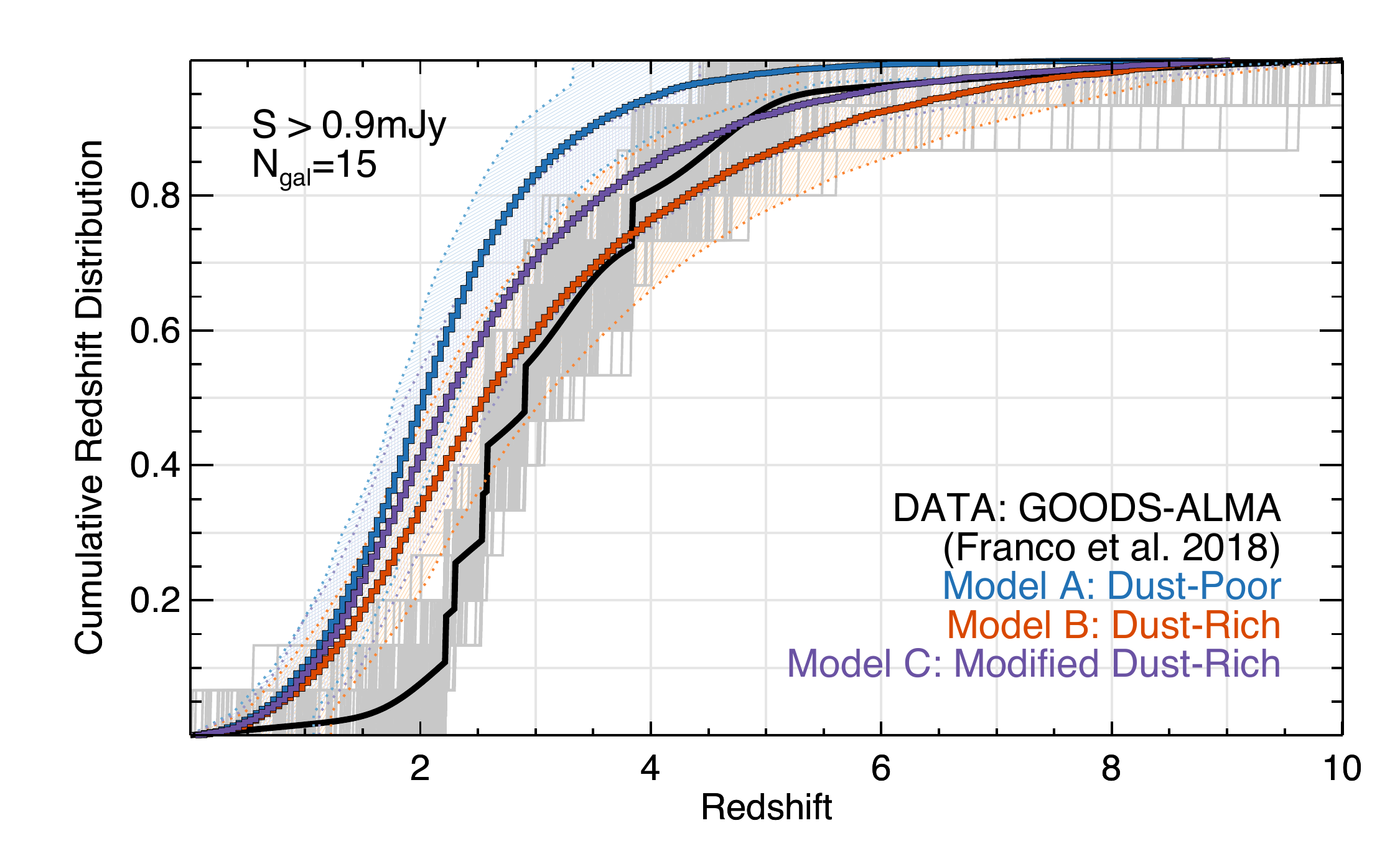}\\
\includegraphics[width=0.99\columnwidth]{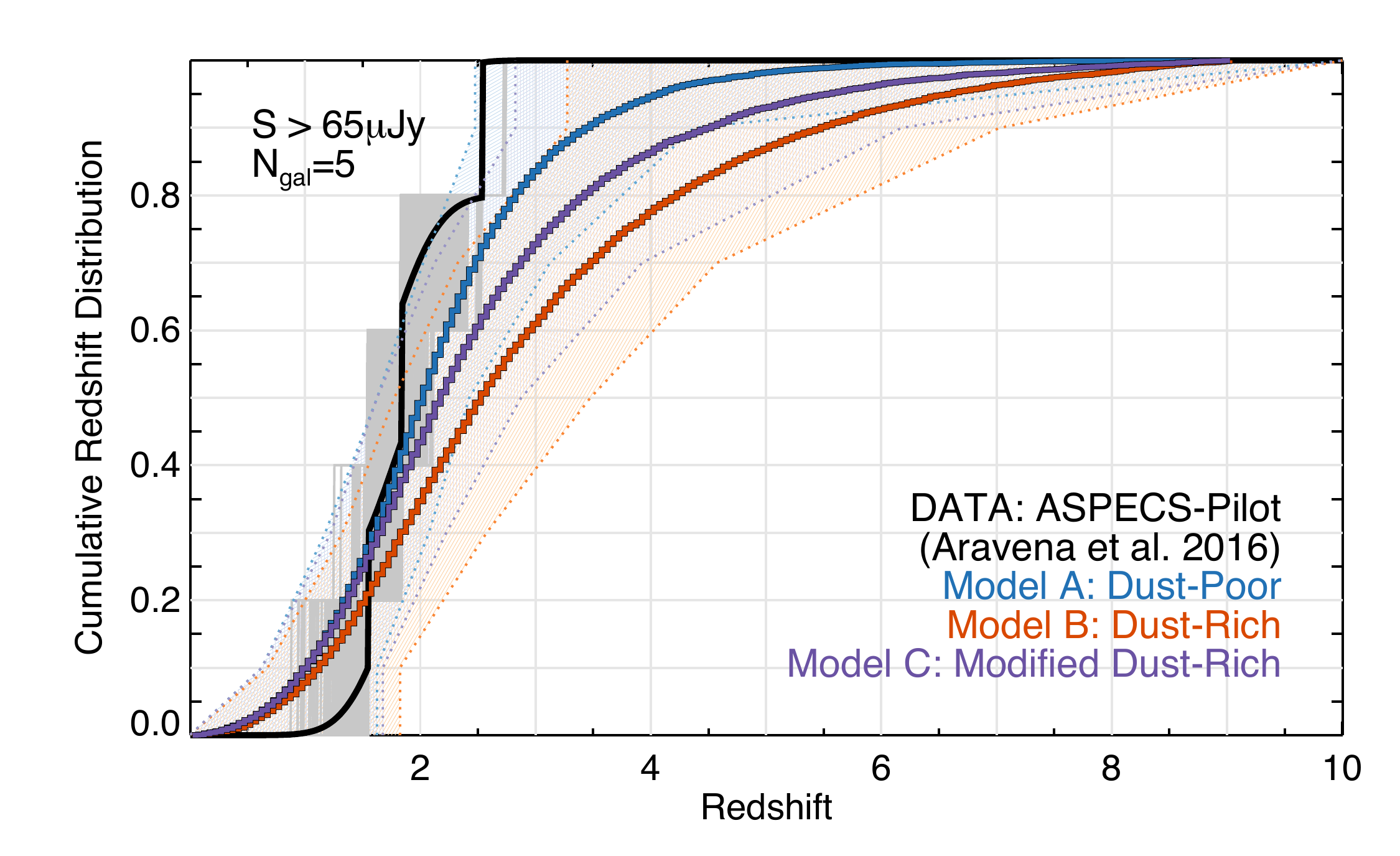}
\includegraphics[width=0.99\columnwidth]{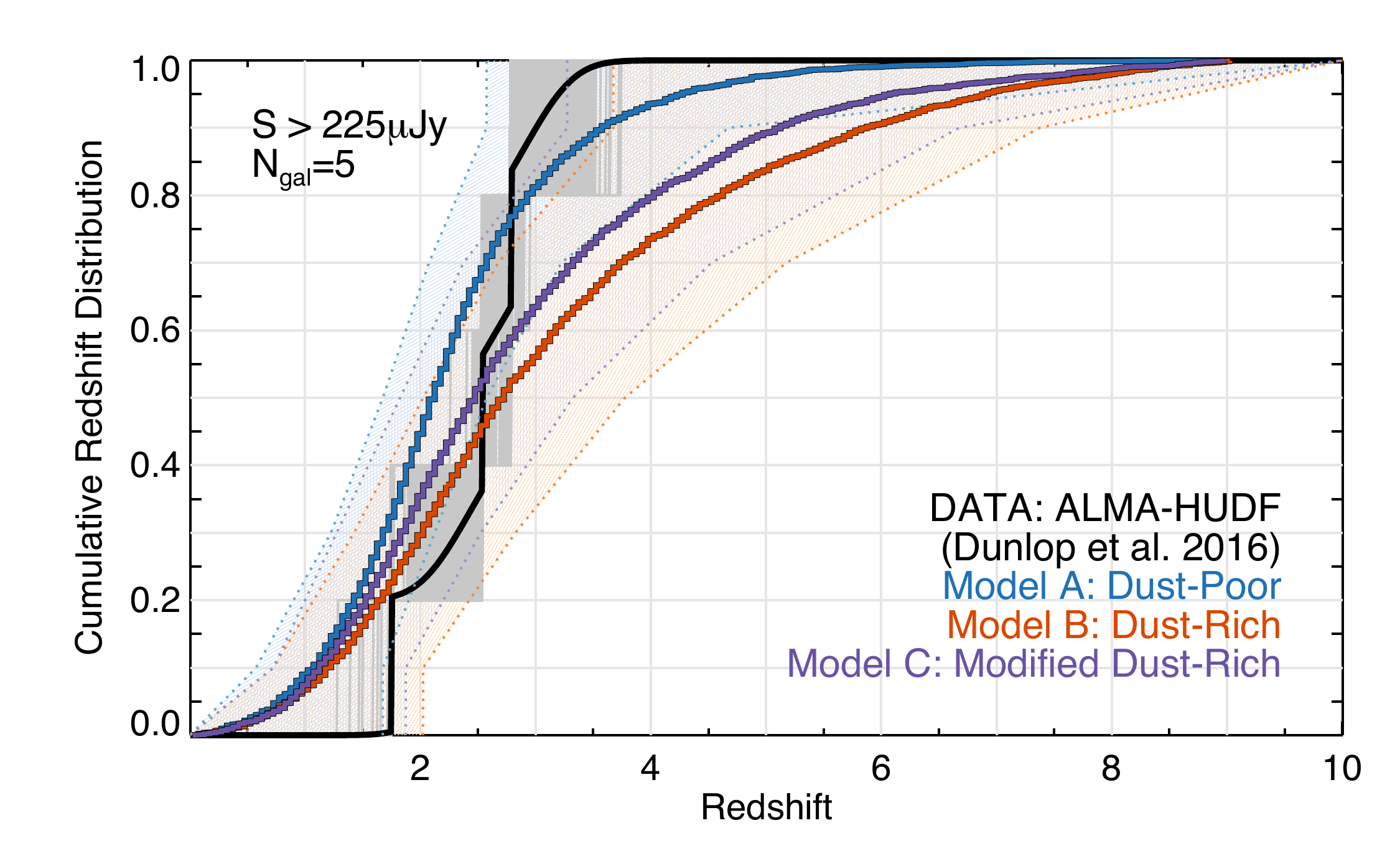}
\caption{Comparison of measured redshift distributions of $>$5$\sigma$
  sources in the 1.2\,mm ASPECS-Pilot sample \citep{aravena16b}, the
  1.3\,mm HUDF sample \citep{dunlop16a} and the 1.2\,mm GOODS-ALMA
  sample \citep{franco18a} against our three models: the dust-poor
  Model A (blue), the dust-rich Model B (orange) and the modified
  dust-rich Model C (purple).  Redshifts are a mix of photometric and
  spectroscopic redshifts.  The gray (data), and light shaded regions
  represent the uncertainty distributions given the sample size of
  ASPECS (5 galaxies), HUDF (5 galaxies), or GOODS-ALMA (15 galaxies)
  in addition to redshift uncertainty for the subsample. Small
  deviations in the faint-end slope of the luminosity function does
  not impact the measured redshift distribution significantly
  (i.e. the difference between $\alpha_{\rm LF}=-0.6$ and $\alpha_{\rm
    LF}=-0.69$ for Model A is indiscernible, though it does impact the
  total number of sources identified above the significance threshold,
  as shown in Figure~\ref{fig:alphafix}).  These samples limited by
  small number statistics are not large enough to distinguish between
  competing models. In addition, the GOODS-ALMA analysis could be
biased against low-redshift sources that are probably larger and
resolved out of the map.  }
\label{fig:nz}
\end{figure*}

As a more robust test and one lending itself to the full information
available in our model, we also estimate the contamination and
completeness of our simulations by comparing the list of injected
sources with the list of detected sources.  For simplicity we assume
galaxies are point sources, unresolved on all spatial scales of our
simulations.  Figure~\ref{fig:cc} shows both completeness and
contamination rates as a function of flux density and signal-to-noise.
Contamination (per bin) is the fraction of sources in the output
catalog at that flux density which lack corresponding input sources
within a beamsize of the source centroid.  Completeness is the number
of sources (per input flux density) that are identified at any
significance $>$3.5$\sigma$ in the output catalog.  It is somewhat
concerning that the expected contamination rate is above 10\%\ below
${\rm SNR}=5$.  As noted also in \citeauthor{dunlop16a}, the high
contamination rate at $3.5<\sigma<5$ (compared to single-dish results)
is likely due to the incredibly high number of independent beams in
ALMA maps.  Therefore we advise future ALMA deep field programs to
consider sources at lower SNR ($3.5\sigma<{\rm SNR}<5\sigma$) cautiously.
It is even a possibility that a positive spike in the ALMA map could
correspond with an OIR counterpart accidentally; we measure this type
of accidental counterpart identification at the level of
$\sim$9\%\ above $F125W<28$ \citep[using the HUDF photometric catalog
  from][]{rafelski15a}.



In contrast to the \citeauthor{aravena16b} and \citeauthor{dunlop16a}
results, the tension between our estimates (85--146) and the
\citet{franco18a} results (20 sources) requires an analysis of angular
resolution.  There is some added complication due to their data
acquisition in an extended baseline mode (achieving a native
resolution of $\sim$0.2--0.3$''$).  This could, in principle, lead to
a lower detection rate because sources larger than these spatial
scales could be resolved out of the mosaic.  To counter this effect,
the authors taper the map to a resolution of 0.6$''$ with the
intention of recovering any missed extended sources in the original
mosaic\footnote{Indeed, of the 20 sources identified in their tapered
  0.6$''$ map, only 14 are found in the higher resolution images.} in
addition to reducing the number of independent beams that cause excess
source contamination at low SNR thresholds.  While the tapered map
does recover some missed sources, \citeauthor{franco18a} then further
discuss the effects of galaxy size on detectability in the tapered
map.  They find a very high completeness for point sources, but a
drastically lower completeness for galaxies of even modestly larger
sizes (with FWHM ranging 0.2--0.9$''$). Using the \citet{hodge16a}
measurements of DSFGs from ALESS as a benchmark, we estimate
$\sim$1\,mJy sources might have typical FWHM sizes of 0.4--0.5$''$,
resulting in 75--95\%\ incompleteness.  Indeed, their estimated
cumulative number counts for sources with $S_{1.1}>0.7$\,mJy gives
61$^{+50}_{-58}$ sources that should be found in the map (contrasting
with the 20 sources found).  This is in-line with our predictions of
85--146 sources from Models A--C.  It is worth reiterating that our
simulations input all galaxies as unresolved point sources.  The
\citeauthor{hodge16a} work highlights that even the 0.6$''$ tapered
map is at risk of resolving sources, leading to further source
incompleteness.  This emphasizes the importance of more compact ALMA
configurations (with larger beamsize) to conduct such blind deep field
surveys.

\subsection{Redshift Distributions}

However uncertain, the rate of false detections is critical to the
interpretation of the 1.2\,mm ALMA-detected redshift distributions
and the answer to the question of why there are so few high-$z$
galaxies detected in ALMA deep fields.  In this section we explore the
predicted and measured redshift distributions for 1.2\,mm samples.  We
first compare against the \citet{aravena16b} and \citet{dunlop16a}
samples, and then follow with a discussion of the \citet{franco18a}
sample.

While both the \citet{aravena16b} and \citet{dunlop16a} analyses
includes sources identified down to a significance of 3.5$\sigma$, our
analysis suggests that 40--80\%\ of sources at that significance are
spurious.  Unfortunately, the existing maps contain very few high
significance sources, and a detection threshold of 5$\sigma$ leaves us
with five sources in each data samples.  One high significance source
is in both the \citet{aravena16b} and \citet{dunlop16a}
samples\footnote{This source is UDF3 in \citeauthor{dunlop16a} and C1
  in \citeauthor{aravena16b} at 03:32:38.53--27:46:34.6, at a redshift
  of $z=2.543$.  Somewhat concerning is the discrepancy between
  reported flux density measurements, with UDF3 reported to have a
  1.3\,mm flux density of 863$\pm$84\,\uJy, while C1 reported to have
  a 1.2\,mm flux density of 553$\pm$14\,\uJy.  Though the frequency of
  observations was not identical between these two programs, the
  1.2\,mm flux density should be either equal to or greater than the
  1.3\,mm flux density, due to the shape of galaxies' SEDs on the
  Rayleigh-Jeans tail of cold dust emission.  One might expect such a
  galaxy at $z=2.5$ to have a flux ratio of $S_{1.2}/S_{1.3}=1.3$,
  though the measured ratio is $S_{1.2}/S_{1.3}=0.64\pm0.10$.},
leaving us with only nine unique sources identified at $>$5$\sigma$.
Nevertheless, including sources found at lower significance could
substantially contaminate the analysis of source redshift
distributions, and so we choose to only compare with the most robust
subset.

The comparison with the \citet{franco18a} work is in some ways more
straightforward, because the sample is larger, but more complex
because there is an additional selection bias folded into the
comparison: we know that galaxies that are more extended in their
millimeter emission are more likely to be excluded from the sample.
The implications of this bias on the redshift distribution are
unclear.  Of the 20 galaxies identified in their map, we compare the
redshift distribution of 15 of those to our models, with flux
densities $S>0.9\,$mJy (representing a 5$\sigma$ detection threshold
with a 0.18\,mJy RMS).  It could be argued that low redshift galaxies
might be physically larger \citep[and thus subtend larger angles,
  despite the roughly constant angular diameter distance beyond
  $z\sim1$;][]{van-der-wel14a}.  Thus they might be preferentially
filtered out due to their size, being extended on spatial scales
$\sim$1$''$.  However, this trend of increased size at lower-redshift
has not been shown conclusively in dust continuum tracers; the best
measurements to-date contain $\sim$20 galaxies \citep{hodge16a} that
also might be impacted by a luminosity and dust-temperature bias.

Despite the small number of sources available for comparison (5, 5,
and 15 in the three nested maps), we can compare the shape of the
cumulative redshift distribution for these unequivocal, reliable
detections with our model output to see if they broadly agree.
Figure~\ref{fig:nz} presents these comparisons.  The comparison is a
bit unfair, given that we are limited to a handful of galaxies in each
sample and our model is representative of thousands of sources
detected over several tens of arcmin$^2$. For this reason, we
illustrate the model uncertainty randomly drawing many subsets of
sample size $n=5$ or $n=15$ from our large simulated sample.  The
shaded regions on Figure~\ref{fig:nz} represent the inner 68$^{th}$
percentile of those subsets. The gray curves represent Monte Carlo
draws of the data from a cumulative redshift probability distribution
for each of the five (or 15) galaxies in each sub-sample,
incorporating errors due to photometric redshifts.

It is clear from these plots and the associated uncertainty that
current datasets are not constraining or able to distinguish between
models.  The comparison with \citet{franco18a} in particular does show
a deficit of low-redshift sources, which could be attributable to the
high angular resolution of the observations, although that is yet
unclear.

If any conclusions can be drawn from this redshift distribution
analysis, it is that the apparent lack of very high-redshift
detections in ALMA deep field pointings to-date are a direct result of
the limiting survey area (the fact that they have only been
pencil-beam surveys), depth, and the intrinsic property of the IRLF
that is known at least out to $z\sim2$ directly, and through this work
more indirectly: that the faint-end slope $\alpha_{\rm LF}$ is
shallow.  As a result of the shallow faint-end slope, the expected
redshift distributions for 1.2\,mm surveys of this size is between
$1.7<\langle z\rangle<3.5$.
Samples of tens to hundreds of $>$5$\sigma$ detected sources are
needed to make distinctions between models (and we discuss possible
observational strategies for doing this in more optimal bands later in
\S~\ref{sec:future}).


\subsection{Analysis of UV-bright Population}\label{sec:bouwens}

An alternate approach to the interpretation of ALMA deep fields is to
analyze expected detection statistics of UV-selected galaxies already
identified in the field of view.  In this section we focus on the
analysis of \citet{bouwens16a}, hereafter B16, who present a sample of
35 Lyman Break Galaxies (LBGs) which they expected to detect in the
ASPECS-Pilot map of \citet{aravena16b} as well as the analysis of
\citet{capak15a} who present 12 rest-frame UV-selected galaxies at
$z\approx5.5$ with dust continuum observations.  Both of these works
argue that high-redshift LBGs might be relatively dust-poor compared
to lower redshift ($z\sim2$) analogues because their ratio of IR-to-UV
luminosity is lower at a given rest-frame UV color.

With only seven sources detected with OIR counterparts in the
ASPECS-Pilot map\footnote{Seven sources are found with OIR
  counterparts in \citet{aravena16a}, although only three of those
  sources overlap with the list of 35 LBGs analyzed in B16. B16 quotes
  that six LBGs are tentatively detected (including those three
  $>$3.5$\sigma$ sources) by pushing the significance threshold down
  to 2$\sigma$.  Four of the seven $>$3.5$\sigma$ detections are not
  included in the B16 LBG samples as they sit at lower redshifts.},
\citeauthor{bouwens16a} conclude that the most reliable way of
estimating dust emission in high-$z$ galaxies is by scaling their
stellar masses, not using their rest-frame UV colors that would have
predicted 35 detections by their calculations.  While other studies
often make use of the IRX-$\beta$ relationship -- the relationship
between the ratio of IR to UV luminosity (IRX$\equiv L_{\rm IR}/L_{\rm
  UV}$) to the rest-frame UV spectral slope, $\beta$ -- to infer dust
luminosity, \citeauthor{bouwens16a} argues that using that method can
dramatically over-predict mm-wave flux densities for individual
rest-frame UV-selected galaxies\footnote{On this point, we agree with
  \citet{bouwens16a} that scaling from IRX-$\beta$ can dramatically
  over-predict mm-wave flux densities, although we caution that the
  SED assumptions made by \citeauthor{bouwens16a} could be improved
  upon further.}, this based on the low rates of detection for LBGs in
the ASPECS-Pilot map.  Adopting a stellar mass predictor for $S_{1.2}$
instead, \citeauthor{bouwens16a} suggest that this can be used across
a range of redshifts (to beyond $z\simgt4$) as long as there is a
monotonic change in galaxy SEDs with redshift, such that they increase
in dust temperature, suggesting $T_{\rm dust}\propto(1+z)^{0.32}$.
The evolution in dust temperature is deemed necessary to account for
lower perceived flux densities on the Rayleigh-Jeans tail of blackbody
emission for high-$z$ galaxies compared to those at low-$z$.

\begin{figure}
\includegraphics[width=0.99\columnwidth]{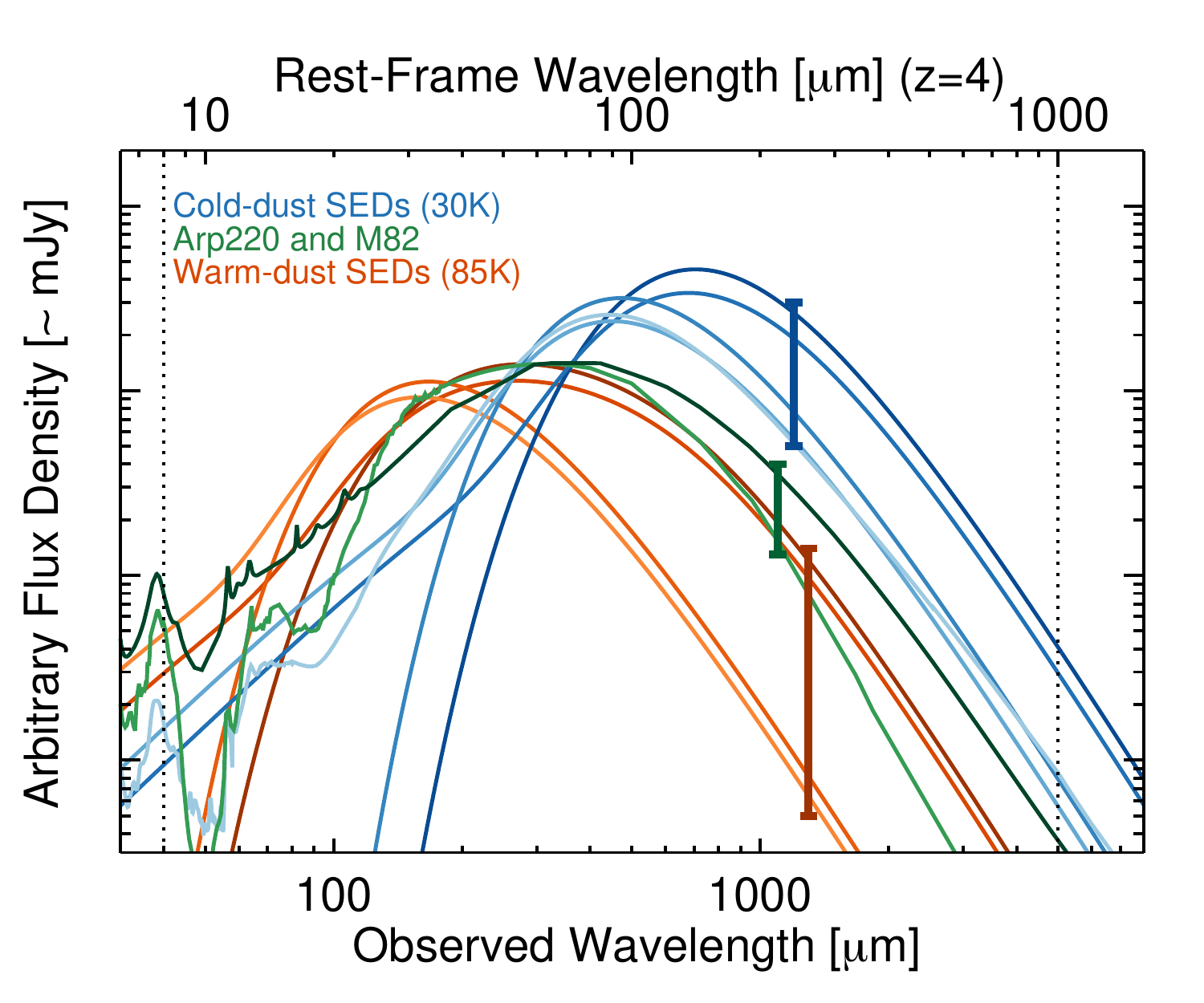}
\caption{Here we show a set of SEDs for a $z=4$ galaxy all with the
  same integrated IR luminosity between 8--1000\um\ in the rest-frame.
  All five blue curves adopt a cold-dust temperature (in this case
  30\,K).  Variation among the cold-dust SEDs is due to: inclusion or
  not of a mid-infrared powerlaw component, and whether or not the SED
  is assumed to be optically thin at all wavelengths (those that are
  peak at shorter rest-frame wavelengths than those assumed to be
  optically thick to $\lambda\approx100$\,\um).  We also include
  comparison SEDs for both Arp\,220 and M82 as local examples of
  galaxies with intrinsically warm dust SEDs (green curves, Arp\,220
  slightly darker of the two).  The warm-dust SEDs shown in the orange
  curves assume a dust temperature of 85\,K; the variation is, again,
  due to opacity assumptions and to a lessor extent, to inclusion of
  an even hotter-dust mid-infrared powerlaw.  Observed 1.1--1.3\,mm
  flux densities for these \lir=$10^{12}$\,\lsun\ SEDs are shown with
  vertical lines, ranging over two decades in flux density.}
\label{fig:seddiff}
\end{figure}
Here we provide an alternate interpretation from B16,
suggesting instead that the functional form of the adopted SED and
reference IRX-$\beta$ relationship matter a great deal to the
interpretation of galaxies' dust luminosities and that no such
evolution in dust temperature is necessary to explain the results.
The most significant differences between our analyses are:
\begin{itemize}
\item Differences in the assumed reference IRX-$\beta$ relationship:
  The derived empirical relationship between IRX and $\beta$ from
  \citet{meurer99a} is offset toward bluer-than-intrinsic colors due
  to differences in aperture sizes of the original
  measurements.\footnote{The {\it IUE} spacecraft measuring the UV
    luminosity and colors of nearby starburst galaxies had a limited
    field of view, only able to image galaxies' cores, while the {\it
      IRAS} far-infrared data used to calculate $L_{\rm IR}$ for the
    same galaxies was unresolved and includes emission on much larger
    scales.  This is discussed extensively in \citet{takeuchi12a} and
    \citet{casey14b}.}  The aperture-corrected calibration of
  IRX-$\beta$, for the exact same sample of local starburst galaxies,
  is given in \citet{takeuchi12a}.  For given measured values of
  $\beta$ and $L_{\rm UV}$, use of the \citeauthor{meurer99a} curve
  will result in a factor of $\sim$0.3\,dex overprediction of $L_{\rm
    IR}$ in comparison with the \citeauthor{takeuchi12a} relation.
  This impacts the inferred $L_{\rm IR}$ values in B16 used to predict
  1.2\,mm flux densities with IRX-$\beta$.  This discrepancy also
  impacts the perceived significance of the disagreement between the
  \citet{capak15a} sample and the `Calzetti' dust attenuation curve,
  although some of that tension was reduced by an updated analysis of
  the rest-frame UV colors in \citet{barisic17a}.
\item Differences in assumed SEDs used to map $L_{\rm IR}$ to
  $S_{1.2}$: B16 explores several types of SEDs but adopts a fiducial
  35\,K modified blackbody SED to scale between 1.2\,mm flux density
  ($S_{1.2}$) and IR luminosity (integrated 8--1000\um) for all
  UV-selected galaxies.  The SEDs we use to map between $S_{1.2}$ and
  L$_{\rm IR}$ differ primarily because they include a mid-infrared
  powerlaw component, which can contribute 10--30\%\ to the total IR
  luminosity of a given galaxy. Physically it comes from much less massive, isolated knots of hot dust
  heated by discrete sources throughout the galaxy, like OB
  associations or an AGN \citep[see C18 and][for details]{casey12a}.
  The differences between a modified blackbody and a modified
  blackbody with a mid-infrared component has an effect such that, for
  a fixed L$_{\rm IR}$, the flux densities on the Rayleigh-Jeans tail
  will be a factor of 0.5--2$\times$ lower for the latter than the
  former.  In other words, SEDs with a mid-infrared component will
  have 1.2\,mm flux densities a factor of 0.10--0.15\,dex lower than
  SEDs without the mid-infrared component at matched $L_{\rm IR}$ and
  SED peak wavelength ($\lambda_{\rm peak}$).  This discrepancy does
  not impact the \citet{capak15a} sample.
\item Difference in assumed SED peak wavelength: we also do not assume
  a single dust temperature (35\,K) for the entire sample of LBGs.  We
  emphasize that the 35\,K B16 modeled SED peaks at a rest-frame
  wavelength of $\sim$85\,\um\ due to the assumption of an
  optically-thin SED, while we would instead predict rest-frame peak
  wavelengths in the range of 100--120\,\um\ for galaxies with SFRs of
  1--10\,\sfr; Figure~\ref{fig:seddiff} illustrates some of the
  dramatic differences in SEDs with the same dust temperature and
  L$_{\rm IR}$ but different opacity models.  For a fixed $L_{\rm
    IR}$, the cooler SED that we assume results in a higher predicted
  flux density by 0.3--0.5\,dex at 1.2\,mm than the warmer SED assumed
  by B16, but with the inclusion of the mid-infrared powerlaw
  component above, the impact of this SED shift is reduced to
  0.15--0.35\,dex.
\end{itemize}
Taking these effects into account and attempting to predict new flux
densities for the same set of 35 LBGs analyzed in B16 (three of which
are detected at $>$3.5$\sigma$), our predictions are a factor of
0.1--0.2\,dex lower than the predictions quoted in B16.  Specifically,
using the \citeauthor{takeuchi12a} scaling and our SED assumptions, we
would predict 15 of 35 sources detectable at $>$3.5$\sigma$.  Using
the SMC attenuation curve \citep{pettini98a} and our SED assumptions,
the detectable number would drop to 10 of 35.
We also test a stellar mass-based predictor of flux density by scaling
stellar mass and measured $L_{\rm UV}$ to $L_{\rm IR}$ using the
empirical relationship between stellar mass and obscured fraction of
star-formation, $f_{\rm obs}$ \citep{whitaker17a}.  Overall, the mass
predictor would estimate 14 detections out of the 35.  While all of
our predictors are off the mark and predict more detections than exist
among this sample of LBGs (though SMC-like dust comes closest), we
note that there is very little correlation overall between either our
predicted flux densities, those of B16, and measured flux densities --
a testament to the relative difficulty of inferring galaxies' dust
content or luminosity from stellar emission alone.

\begin{figure*}
\centering
\includegraphics[width=1.6\columnwidth]{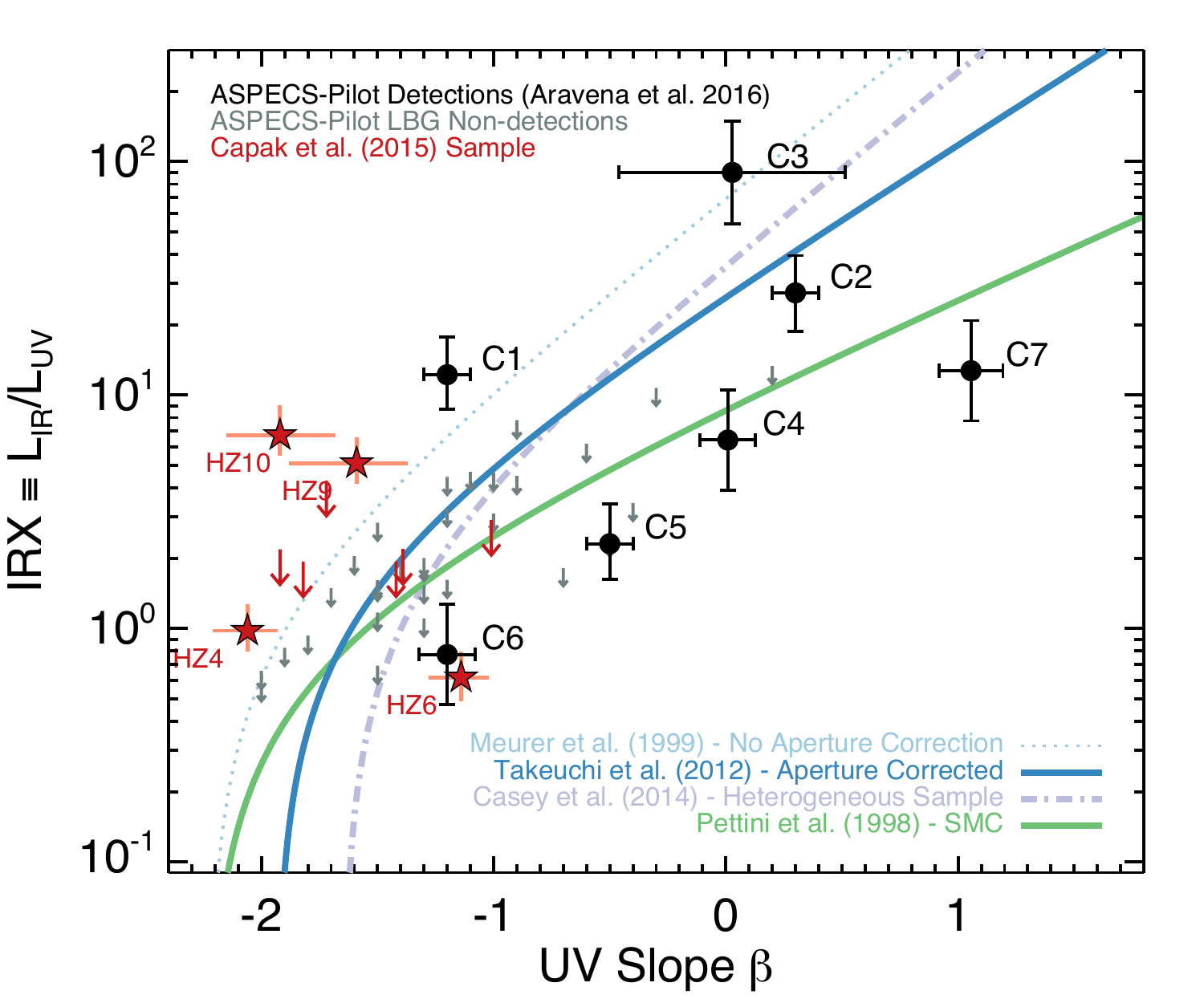}
\caption{The IRX-$\beta$ relationship for the LBG galaxies described
  in B16.  The seven detections that have OIR counterparts
  \citep{aravena16b} are shown as black points.  LBGs in the
  ASPECS-Pilot map without direct detections are shown as gray upper
  limit arrows.  The $z\sim5.5$ galaxies from \citet{capak15a} are
  also shown in red with updated $\beta$ values from
  \citet{barisic17a}; those with dust continuum detections are stars,
  while upper limits are arrows. The scatter of data points here about
  the \citet{takeuchi12a} IRX-$\beta$ relation (thick blue line) for
  detected sources is representative of intrinsic scatter in galaxy
  populations based on dust geometry; the fact that more skew below
  the IRX-$\beta$ relationship perhaps indicates more consistency with
  and SMC-type curve (green line).  The upper limits given by
  non-detections cannot rule out either SMC or Milky Way-type dust,
  which is inconsistent with the findings of B16 that claim high-$z$
  sources fall below the SMC curve with 95\%\ confidence.  We
  attribute the difference in conclusions to the adopted form of the
  far-infrared SED in addition to the difference between the
  \citeauthor{meurer99a} and \citeauthor{takeuchi12a} curves.  We also
  overplot the \citet{meurer99a} and \citet{casey14b} curves as dotted
  light blue and dot-dashed lavender, respectively.}
\label{fig:irxb}
\end{figure*}

Figure~\ref{fig:irxb} shows the inferred IRX-$\beta$ relationship for
the detections (and non-detections) in the ASPECS-Pilot map, in
addition to the \citet{capak15a} sample, with revised UV colors from
\citet{barisic17a}.  Here, IRX (or the limit thereof) is re-derived
for each galaxy in the sample using the observed flux density (or
limit thereof) and a family of SEDs deemed most appropriate for the
source given its $S_{1.2}$.  As discussed in C18, we observe that
$L_{\rm IR}$ relates directly to $\lambda_{\rm peak}$ with some
scatter.  To predict a IR luminosity from flux density, we generate a
family of SEDs (all with a mid-infrared powerlaw component included)
that mirror the observed scatter in $L_{\rm IR}-\lambda_{\rm peak}$.
We then search for all possible SEDs that have the observed flux
density at 1.2\,mm and use it to generate a probability density
distribution in $L_{\rm IR}$.  Then using the measured $L_{\rm UV}$
(with associated uncertainty), we infer IRX and a realistic
uncertainty from the single flux density measurement.

Of the seven ASPECS-Pilot sources detected with OIR counterparts, five
sit below the canonical IRX-$\beta$ relationship described by
\citet{takeuchi12a} for blue, compact starbursts, while two are
significantly offset above the relation.  Non-detections are shown as
3.5$\sigma$ upper limits.  Unlike the results of B16, our results
suggest that the vast majority of these upper limits (29/32) are
consistent with the Calzetti dust attenuation law.  The difference in
conclusions is due both to differences in modeled SEDs and reference
Calzetti IRX-$\beta$ relationships.

The IR luminosities of the $z\sim5.5$ LBG sample \citep{capak15a} were
fit very similarly to the SEDs in this paper, although lacking the
luminosity dependence of $\lambda_{\rm peak}$.  Note that in this
paper we re-derive $L_{\rm IR}$ for the sample using the same method
used for the ASPECS-Pilot sample\footnote{Instead of adopting a fixed
  range of dust temperatures irrespective of IR luminosity, here we
  adopt the observed $L_{\rm IR}-\lambda_{\rm peak}$ relationship
  shown in C18.  The scatter in SEDs is similar to the original
  assumptions of \citet{capak15a}.}.  It is worth noting that the
plotted upper limits on IRX in both \citet{capak15a} and
\citet{barisic17a} are 1$\sigma$ limits; in Figure~\ref{fig:irxb} we
have shown more conservative 3.5$\sigma$ upper limits.  Combined with
the shift toward bluer colors as measured by improved rest-frame UV
imaging in \citet{barisic17a}, the more conservative IRX upper limits,
and the comparison to the \citeauthor{takeuchi12a} curve instead of
\citeauthor{meurer99a}, the relative tension between the Calzetti dust
attenuation law and the $z\sim5.5$ sample is significantly reduced.

The important finding here -- for both the \citeauthor{capak15a} the
\citeauthor{bouwens16a} high-$z$ samples -- is that this consistency
with the Calzetti dust attenuation law cannot be directly ruled out
from existing measurements, even with typical SED assumptions that
hold for much lower redshift galaxies. 

Other works \citep[e.g.][]{faisst17a} have argued that there is
significant tension between measurements and the Calzetti dust
attenuation law for such cold SEDs, and that only much warmer-dust
SEDs $\ge$60\,K could ease the tension (whereby warmer-dust SEDs have
much lower $S_{1.2}$ for a given $L_{\rm IR}$ than colder-dust SEDs).
%
%
\citet{faisst17a} draws on the characteristics of three local galaxy
analogues selected from {\it GALEX} samples as Lyman-$\alpha$
emitters, where all three galaxies have warm dust SEDs with steep Wien
tails. In other words, their SEDs are more homogeneously represented
by a single luminosity-weighted temperature than a powerlaw
distribution of temperatures found in the ISM of typical massive
galaxies.  We illustrate the difference between our SED assumption and
the warmer-dust SEDs in Figure~\ref{fig:seddiff}; this highlights how
a diverse range of SEDs with the same integrated \lir\ might result in
dramatically different measured flux densities on the Rayleigh-Jeans
tail.

While our results do not require such hot temperatures to ease tension
between measurements and local IRX-$\beta$ relationships (Calzetti or
SMC), we do not wish to completely dismiss the idea that high-$z$
galaxies might have much hotter dust.  Indeed, this claim does have
some grounding in physical arguments as discussed in
\citet{behrens18a} who present the results of a hydrodynamic zoom-in
simulation of a $z\sim8$ galaxy named Althaea whose
luminosity-weighted dust temperature is very warm (91\,K, peaking at a
rest-frame wavelength of 50\um), and exhibits a sharp Wien cutoff.
They argue that deeply embedded young star clusters might irradiate
compact regions of early galaxies' ISM, such that the strong
interstellar radiation field leads to much warmer intrinsic
temperatures than are seen for more mature galaxies whose ISM might be
predominantly more diffuse.  They too argue that a Calzetti dust
attenuation law can explain the observed characteristics of galaxies
like A2744\_YD4 at $z=8.38$ \citep{laporte17a}, even if the geometry
of the dust in such nascent systems is distributed quite
differently; but they do so by arguing that this dust is likely much
hotter than $\sim$30\,K.

Though such hot dust is physically plausible in high-$z$ galaxies, our
results demonstrate that it is not {\it needed} to explain the
observed dust characteristics of high-$z$ LBG samples.  Either warmer
temperatures and a steep Wien tail fall-off or a cool temperature and
a mid-infrared powerlaw component consistent with lower-redshift
galaxies ($\alpha_{\rm MIR}\approx2$) can rectify the perceived global
offset from the expected IRX-$\beta$ relationship.


\begin{figure}
\includegraphics[width=0.99\columnwidth]{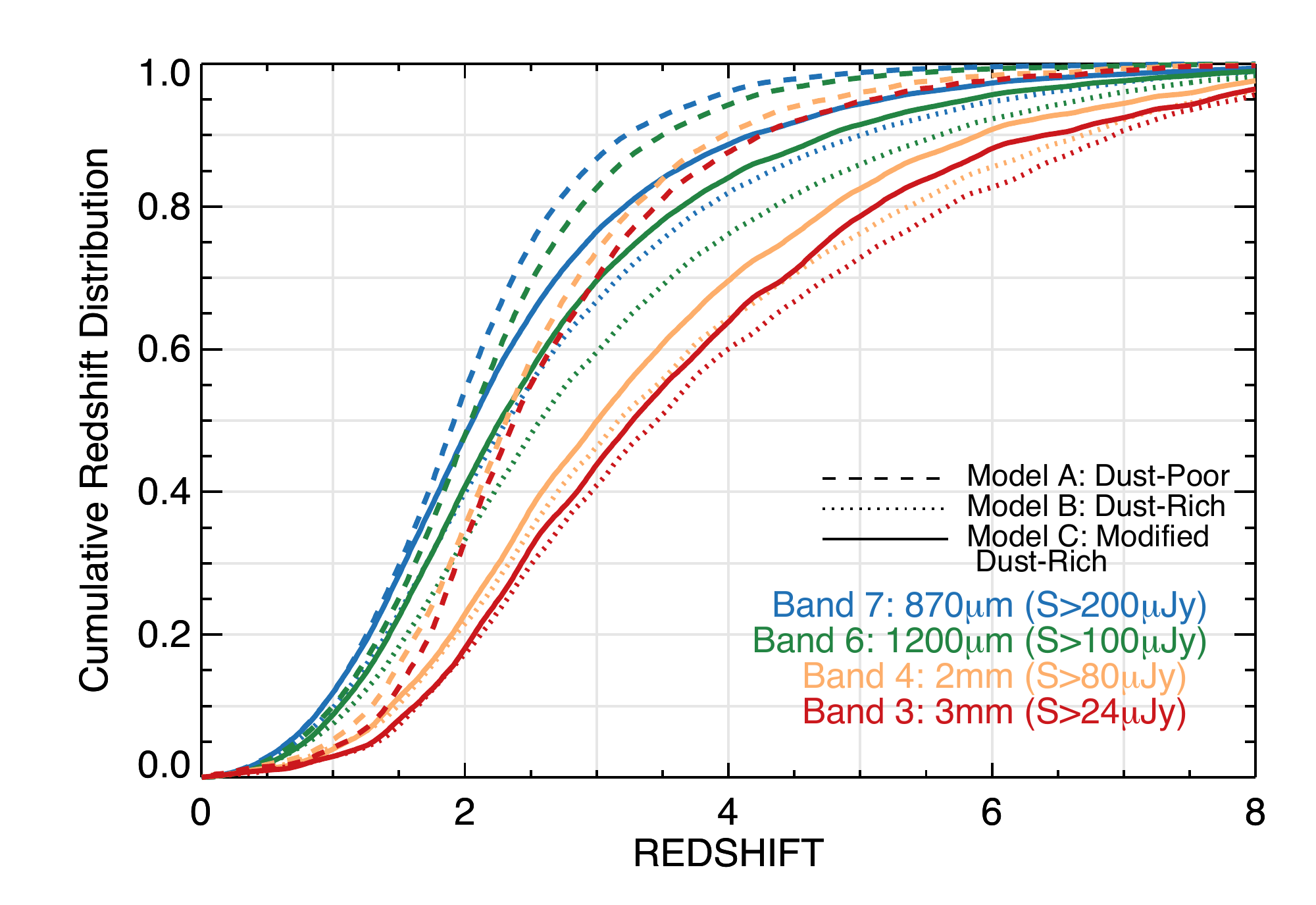}
\caption{The predicted cumulative redshift distributions for sources
  identified at $>$5$\sigma$ in ALMA deep fields conducted at
  870\um\ (Band 7; blue), 1.2\,mm (Band 6; green), 2.0\,mm (Band 4;
  peach), and 3.0\,mm (Band 3; red).  The 1$\sigma$ RMS assumed is for
  this figure is given in Table~\ref{tab:setup}. Predictions from Model
  A are shown as a dashed-line, while Model B is dotted and Model C is
  solid.}
\label{fig:zdist}
\end{figure}

\section{Outlook for Future ALMA Datasets}\label{sec:future}

The current 1.2\,mm ALMA deep field datasets have only begun to
scratch the surface of possible blank-field ALMA constraints.  While
perhaps some models and predictions would have expected many more
sources in 1.2\,mm maps than exist -- either as a reflection of the
steepness of the faint-end of the UV luminosity function, or the
potentially dust-rich Universe that might have been -- our work
suggests that what has been found so far is perfectly consistent with
expectation from brighter source, single-dish surveys.  This does not
mean to imply that the Universe is {\it less} dusty than previously
thought, nor does it mean that there is a measured statistical absence
of dusty galaxies where they should have been.  In fact, even the most
extreme assumptions of the prevalence of DSFGs at high-$z$, assuming
they dominate all of cosmic star-formation at $z>4$ by over a factor
of 10 (i.e. Model B), cannot be ruled out.  The fact is that 1.2\,mm
pencil-beam surveys do not place a good constraint on the relative
prevalence of dusty galaxies across a range of redshifts by the very
nature of their design.

\begin{figure*}
\centering
\includegraphics[width=0.99\columnwidth]{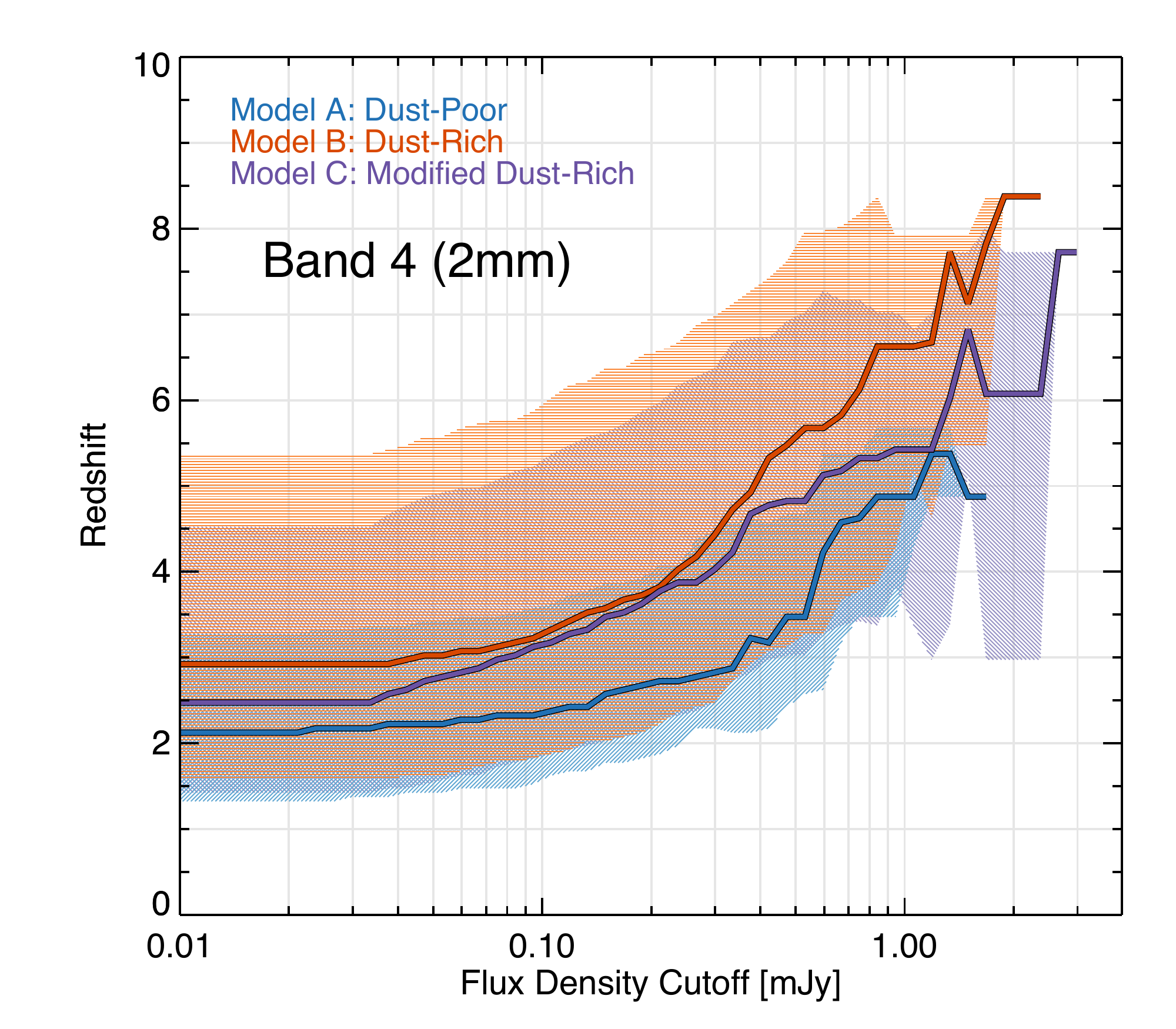}
\includegraphics[width=0.99\columnwidth]{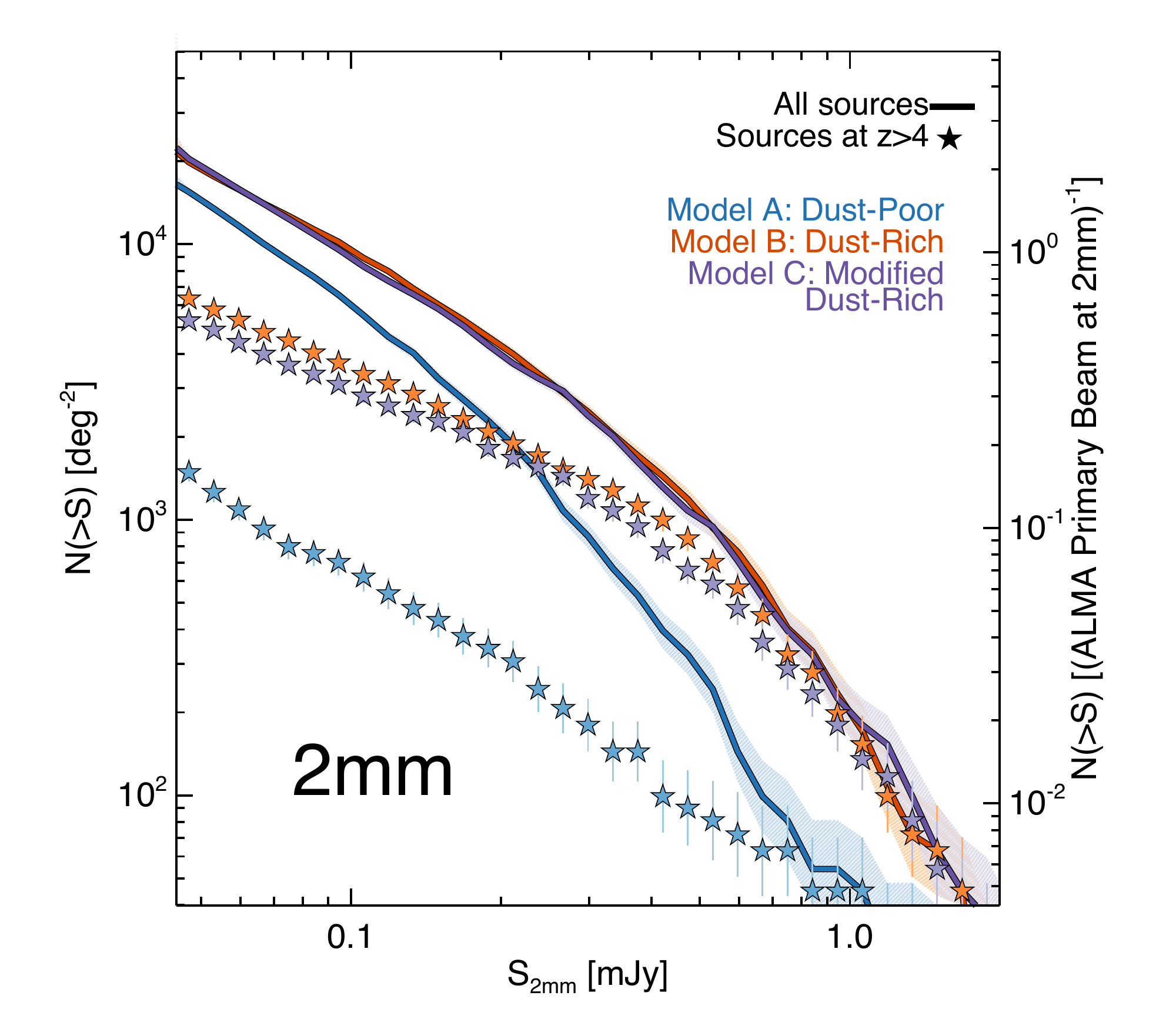}
\caption{Left: The median redshift of our different models (A, B, and
  C) as a function of cutoff flux density at 2\,mm.  The shaded region
  enclosed in the inner 68\%\ of the samples.  This figure shows that
  constraining 2\,mm redshift distributions would be able to
  distinguish between, e.g. model A (dust-poor) and B and C
  (dust-rich) with modest sample sizes, 20--100 galaxies.  The flux
  density regime most sensitive to high-redshifts is 0.2--1.0\,mJy.
  Right: the expected number counts at 2\,mm in our three different
  models (solid lines).  For each model, we give the number counts of
  sources above $z>4$ (stars).  Note that models B \&\ C predict that
  nearly {\it all} of the 2\,mm number counts above 0.5--0.6\,mJy
  should lie at $z>4$. }
\label{fig:2mm}
\end{figure*}

The reason this uncertainty still plagues our efforts to characterize
obscured star-formation in the early Universe is because our community
has focused on the design of ALMA deep fields much the way the
UV/optical community focused and designed deep fields for the {\it
  Hubble Space Telescope}.  The HDF, HUDF, and HFF
\citep{williams96a,beckwith06a,lotz17a} have been extremely rich
legacy datasets purely because the galaxy number density is so high,
even out to $z\sim4-5$, with non-negligible samples out to $z\sim10$.
The high number density is due directly to the slope of the faint-end
of the UVLF, evolving from $\alpha_{\rm LF}^{\rm UV}\approx-1.5$ to
$-2.5$ from $4<z<10$ \citep{finkelstein16a}.  The IRLF by contrast has
a much shallower faint-end slope, reflecting the fact that galaxies do
not become significantly dust-obscured until they are sufficiently
massive \citep[e.g.][]{whitaker17a}.  
Independent evidence to the shallowness of this faint-end slope comes
from the low number density of dusty systems identified behind
clusters, where gravitational lensing can significantly boost sources
flux densities \citep{rawle16a,gonzalez-lopez17a}.
Such a shallow slope, and a
relative dearth of very low-mass obscured galaxies, implies that the
most fruitful mm-wavelength surveys of galaxies are not the same deep,
pencil-beam survey approach that brought us the {\it Hubble} deep
fields.

To evaluate which alternative strategies might be fruitful first
requires a re-assessment of the community's primary science goals.
Because ALMA deep fields lack the remarkable source density of the
rest-frame UV/optical emission of the {\it Hubble} (and eventually
{\it JWST}) deep fields, we are unfortunately not able to answer a
diverse range of scientific questions with a single data product.

\subsection{Needles in the Haystack: Going Deeper does not Reach Farther}

The focus of this paper, and C18, is the search for and census of
dust-obscured galaxies out to very high-redshifts.  What is the
intrinsic shape of the IRLF and how do obscured galaxies contribute to
the overall star-formation rate density of the Universe?  Does the
prevalence of DSFGs at high-$z$ provide any useful constraints on the
growth of massive galaxies within the first 1-2\,Gyr after the Big
Bang?  Answering these questions in particular requires a systematic
follow-up of high-$z$ dust-continuum detected galaxies.

As described in \S~\ref{sec:model} we extend the C18 model into the
ALMA depth, sensitivity and resolution regime by also simulating
blank-fields from 870\um--3\,mm.  Following the same procedure as in
\S~\ref{sec:existing}, to compare with 1.2\,mm existing surveys, we
model the redshift distributions for 870\um\ (band 7), 1.2\,mm (band
6), 2\,mm (band 4) and 3\,mm (band 3) for each of the three models (A,
B, and C) in Figure~\ref{fig:zdist}.  The optimum best-fit values of
$\alpha_{\rm LF}$ are used for Models A and B (--0.69 and --0.49,
respectively), and the depths given represent the conservative
5$\sigma$ cut as was deemed necessary in \S~\ref{sec:existing} to
avoid high rates of false positives.  Overall, Model A (the dust-poor
model) skews towards lower redshifts, Model C is at higher
redshifts, and Model B skews towards slightly higher redshifts yet.
Following the pattern seen for brighter sources in C18, shorter
wavelength surveys selects sources at lower redshifts.  The median
redshifts for Band 7 selected sources is $1.9<\langle z_{\rm
  870}\rangle<2.5$, while Band 6 ranges from $2.0<\langle z_{\rm
  1.2mm}\rangle<2.7$, Band 4 spans $2.3<\langle z_{\rm
  2mm}\rangle<3.2$, and Band 3 spans $2.4<\langle z_{\rm
  3mm}\rangle<3.6$.  

A natural question that follows is whether pushing these surveys
deeper would result in more high-redshift detections, across any or
all of these bands.  For example, the ASPECS project has pushed the
depth of the HUDF map to ASPECS-Pilot depth across the full
4.5\,arcmin$^2$.  Our model predicts between 60--70 sources detected
above $>$5$\sigma$ significance (with 12\,\uJy/beam RMS at 1.2\,mm), a
median redshift between $\langle z\rangle=2.0-2.6$ and between 2--9
sources at $z>4$.  Similarly deep 870\um\ surveys (where
RMS$_{870}\approx 2\times$RMS$_{1.2}$) produce nearly identical
samples, though even fewer detections at $z>4$ due to the slightly
less advantageous negative K-correction.  While it is true that the
larger statistical samples that will come with such surveys will be a
great help in characterizing higher redshift obscured galaxies, the
$z>4$ sources will truly be needles in the haystack: their
spectroscopic identification will likely be exceedingly difficult, and
it will not be immediately clear which sources are at $z>4$ as opposed
to $2<z<4$.

Pushing even deeper, to $\sim$\uJy\ flux densities, what might we
expect to find? At these depths, it is still true that only a minority
of sources, $\sim$5-20\%\ no matter the adopted model, will sit at the
highest redshifts.  In a 1\,arcmin$^2$ survey at 1.2\,mm to
1\,\uJy\ RMS, our model A suggests detection of $\sim$130 sources, 9
of which would sit at $z>4$.  Model B suggests detection of detection
of $\sim$190 sources, 40 of which would sit at $z>4$.  The modeled
flux densities of the highest-redshift sources span the whole range,
5\uJy\ up to $\sim$2\,mJy, similar to the low redshift sources in the
mock map.  Though this \uJy\ regime is potentially fruitful and insightful
to dust emission mechanisms across a range of redshifts, the time
investment required to map such areas is prohibitively large by
today's standards: requiring almost two weeks of on-source time for a
single 1.2\,mm, 0.1\,arcmin$^2$ pointing. Furthermore, additional
precautions are necessary for such observations given the anticipated
dynamic range of sources that are likely to exceed the standard factor
of $\sim$100 between the brightest source in the map and the very deep
target RMS.

Taking a step back and returning to Figure~\ref{fig:zdist}, it is
clear that some surveys are going to be more efficient avenues for
characterizing high-redshift sources than others. In particular,
surveys in Band 4 and Band 3 would provide much higher redshift
samples by effectively filtering out low redshift interlopers.  This
filtering will greatly simplify the process of identifying and
characterizing the highest redshift sources, as a much larger fraction
of the identified samples will sit at $z>4$ (jumping from 2-10\%\ at
$\sim$1\,mm to 30--60\%\ at 2\,mm or 3\,mm, depending on
depth). Whether or not a survey is conducted at 2\,mm or 3\,mm depends
very much on the studies' more precise goals.

\subsection{A Case for 2\,mm ALMA Surveys}

In C18, we advocate for 2\,mm single-dish surveys as an ideal tool for
taking census of dust-obscured galaxies beyond $z\sim4$.  Indeed, that
regime is the most sensitive to extremely luminous (and rare)
starbursts that might have formed the first massive galaxies less than
1\,Gyr after the Big Bang.  In this paper, we shift focus to
 slightly fainter flux densities, smaller area surveys
achievable by ALMA.  Figure~\ref{fig:2mm} shows the anticipated
average redshift of a 2\,mm-selected sample as a function of flux
density cutoff, and also the number density of sources expected for
each model as a function of flux density.  Mirroring the predictions
of Figure~12 in C18, here we see that brighter 2\,mm sources are those
that are expected to sit at the highest redshifts.  Given this
prediction, is there any value in pursuing 2\,mm surveys with ALMA
instead of single-dish facilities like the IRAM 30\,m, JCMT, or the
LMT?

%

We conclude that there is significant value in an ALMA mapping of the
sky in Band 4 for a few reasons.  Currently, there is no immediate
plan to carry out a large-field 2\,mm survey with a single-dish
facility to $\sim$0.1\,mJy sensitivity.  Given the imminent launch of
the {\it James Webb Space Telescope (JWST)} and its goal of studying
galaxy formation and evolution towards very high redshifts, it would
be wise for the broad community to have such a 2\,mm map in hand as
soon as possible to maximize 2\,mm source {\it JWST} follow-up
strategy.

A second reason such a blind-field band 4 mapping would be valuable is
due to the ease with which multiwavelength counterpart identification
and characterization can be carried out, in contrast to the difficult
work of counterpart identification for large beamsizes of single-dish
work.  It is particularly true for high-redshift DSFGs that it will be
challenging to identify multiwavelength characteristics because they
will lack the radio or bright mid-infrared counterparts often used for
cross-band matching
\citep{roseboom10a,roseboom12a,magdis11a,casey12b,casey12c} due to the
contrasting {\it K}-corrections between those wavelength regimes
(where it is positive) and the millimeter (where it is very negative).
A top priority of any such survey will be swift source follow-up to
determine redshifts; knowing the sources' position precisely allows
for a diverse range of multiwavelength follow-up, from the optical and
near-infrared through radio.  
While single-dish 2\,mm maps will cover substantially more area
(though requiring significant time allocations on such single-dish
facilities), it is probable that redshift confirmation will need to
rely on facilities like ALMA for detection of CO or CII, and that
fewer options for follow-up will be a result of lack of precision on
sources' positions.

Though single-dish 2\,mm surveys will always be able to cover much
larger areas of the sky than ALMA, we find that a Band 4 map with
$\sim$1-2$''$ beamsize, 0.08\,mJy/beam RMS, and an area
$\sim$230\,arcmin$^2$ matched to the area coverage of deep OIR surveys
like CANDELS would result in 20 (Model A) to 120 detections (Models B
\&\ C), providing sufficient statistics to distinguish between these
broad models, or favor a new model somewhere between these extremes.
Such a survey is possible with 50\,hours investment in ALMA
time, and in band 4, would not risk resolving out emission on spatial
scales $<$1$''$; although this time investment is significant, it
would be possible to carry out during non-optimal weather conditions
on the Chajnantor Plateau and therefore would be fairly easy to
complete.

\subsection{3\,mm Dust Continuum as a Unique Tool}

Nominally, ALMA Band 3 (3\,mm) would not be an efficient band to
search for dust continuum emitters because galaxies' dust emission is
significantly fainter at 3\,mm than at 2\,mm due to the flux density
fall off on the Rayleigh-Jeans side of the cold dust blackbody.
Indeed, galaxies at $z\sim2$ have an average flux ratio of
$S_{3mm}/S_{2mm}=0.22\pm0.02$ and galaxies at $z\sim5$ have
$S_{3mm}/S_{2mm}=0.28\pm0.04$.  The contrast with 1.2\,mm flux
densities is even more extreme, with $S_{3mm}/S_{1.2mm}=0.04\pm0.02$
and $S_{3mm}/S_{1.2mm}=0.09\pm0.03$ at $z\sim2$ and $z\sim5$,
respectively.  Given such low flux densities at 3\,mm, and the fact
that 3\,mm continuum will be more impacted by CMB heating than shorter
wavelengths, observations must be significantly more sensitive than
surveys at 2\,mm.

\begin{figure*}
\centering
\includegraphics[width=0.99\columnwidth]{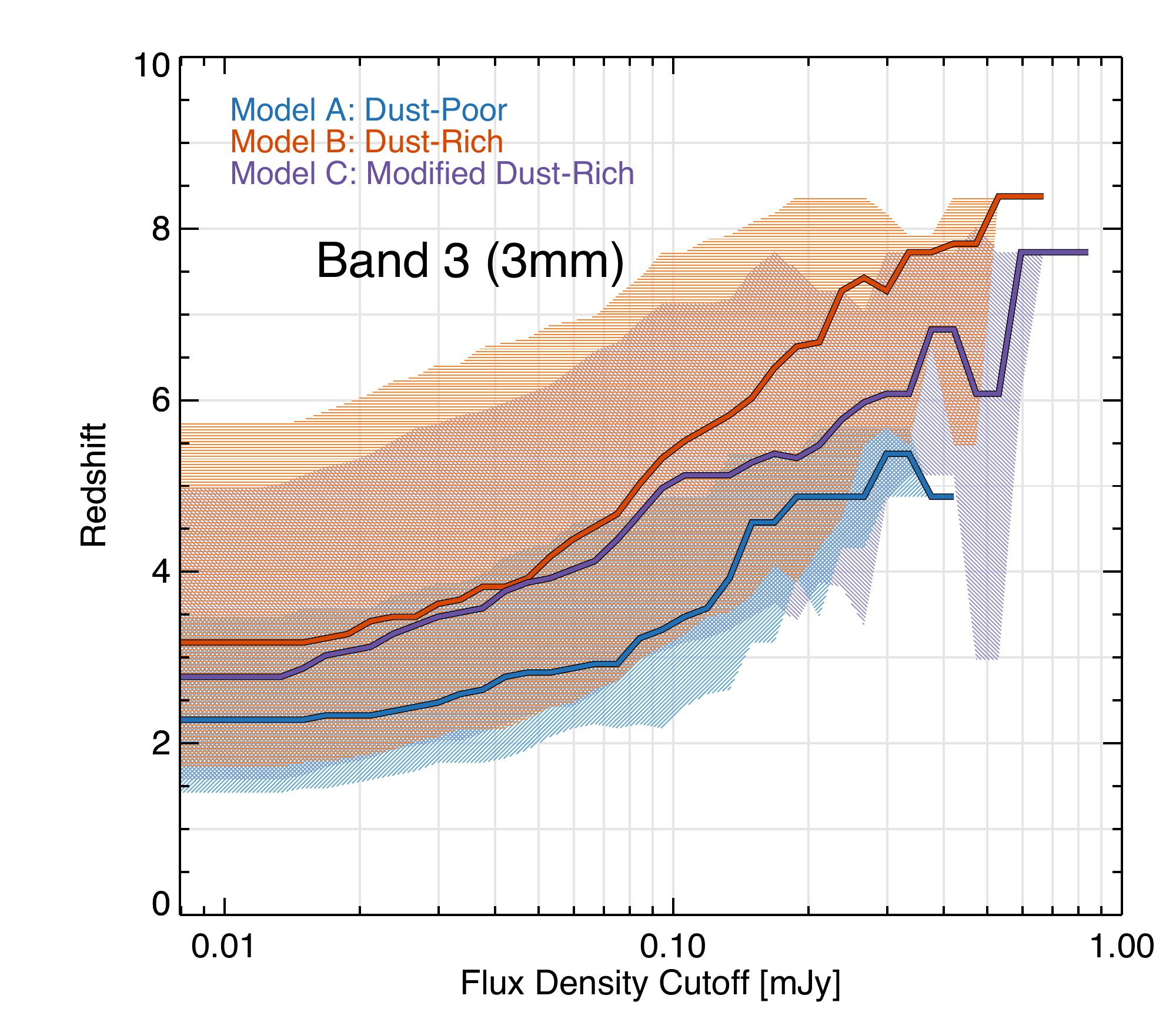}
\includegraphics[width=0.99\columnwidth]{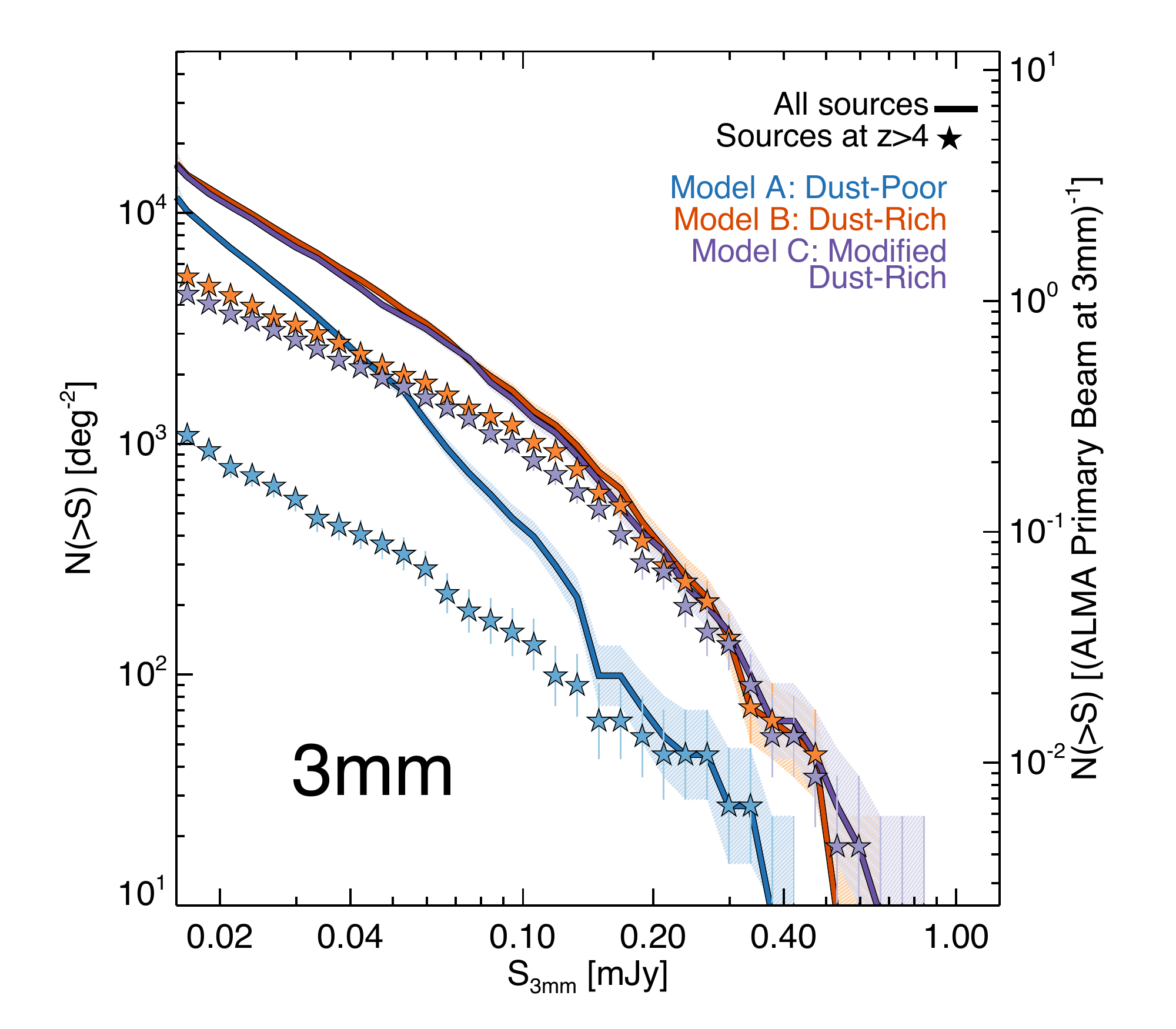}
\caption{This figure is identical to Figure~\ref{fig:2mm} but provides
  the median redshift and cumulative number counts for 3\,mm surveys.}
\label{fig:3mm}
\end{figure*}

However, the lion's share of ALMA Band 3 extragalactic observations
target low-J transitions of CO to study intermediate redshift
galaxies' molecular gas content.  CO(2-1) is accessible in Band 3 at
$1.0<z<1.7$, while CO(3-2) is accessible in Band 3 from $2.0<z<3.1$.
Because such observations require deep sensitivity across frequency
channels $\sim$10\,MHz wide (with the goal of detecting an emission
feature peaking at a few mJy across a few hundred km/s), they are
extraordinarily sensitive in continuum across the 8\,GHz total
bandwidth.  \citet{aravena16b} highlight the depth of 3\,mm continuum
observations in the ASPECS-Pilot project, achieving a continuum RMS of
3.8\,\uJy/beam.  This depth was achieved primarily to detect
transitions of low-J CO blindly
\citep{walter16a,decarli16a,decarli16b}, but it is also sufficiently
deep to detect dust continuum\footnote{Note that flat-spectrum radio
  sources can also generate emission at 3\,mm, although such sources
  are much more rare by number than dusty galaxies.  For example,
  scaling from the 4.8\,GHz number counts of \citet{tucci11a}, we
  estimate a source density of 10$^{-7}$\,deg$^{-2}$ for such
  synchrotron sources.}.  \citet{aravena16b} present one source, C1
(at $z=2.543$), that is detected in their 0.79\,arcmin$^2$ Band 3
pointing.  What more might we expect to see in 3\,mm dust continuum
maps?

Figure~\ref{fig:3mm} shows what our three models predict for the
median redshift of and number counts of 3\,mm continuum-detected
sources, following the format of Figure~\ref{fig:2mm}.  Though 3\,mm
flux densities for matched sources are $\sim$1/5 of the flux densities
at 2\,mm, the redshifts are slightly higher (given the extreme
negative {\it K}-correction at 3\,mm, which is even more steep than
2\,mm).  Like 2\,mm, 3\,mm mapping has the potential for
distinguishing between extreme models like Model A and Model B.  At
the depth of the ASPECS-Pilot project ($S>0.02$\,mJy), Model A
predicts one source per ALMA primary beam, while Models B and C
predict three (in both cases they are thought to sit at $z<4$ more
likely than at higher redshifts given the depth).  Both are
statistically consistent with the measured one source found.

Though 3\,mm mapping is prohibitive for the purposes of finding and
detecting dust continuum emitters, its dual purpose of following-up
lower redshift sources in molecular gas observations render Band 3
observations as a unique tool for constraining the high-redshift IRLF,
as much of the data have already been taken.  For example, with
$\sim$100 Band 3 pointings at 3\,mm (covering an effective area of
$\sim$90\,arcmin$^2$) we can begin to constrain the 3\,mm number
counts and hone in on constraints for the IRLF (J. Zavala \etal, in
preparation).

\section{Conclusions}\label{sec:conclusions}

This paper has extended the backward evolution model from
\citet{casey18a} into the regime of ALMA, allowing deeper and higher
resolution observations than single-dish facilities.  The purpose of
our analysis has been to synthesize existing measurements of
1.1--1.3\,mm ALMA deep fields with submm/mm single-dish surveys, which
generally detect more intrinsically luminous DSFGs.  A simple
extension of the C18 models appears to reproduce the 1.2\,mm ALMA
number counts well, although measurement errors for data samples are
quite large.  We use the source density of \citet{dunlop16a} to refine
our model estimates of the faint-end slope of the IRLF at
$z\simlt2.5$, and devise a third model which is a variant on the
dust-rich model, Model C, where the faint-end slope, $\alpha_{\rm
  LF}$, evolves to become shallower with increasing redshift.  With
larger 1.2\,mm surveys combined with the progress of single-dish
surveys, we may soon be able to refine the measurement of the
faint-end slope of the IRLF out to $z\sim3$.

We find 1.2\,mm ALMA deep fields completely unconstraining for the
high-redshift IRLF.  Because the faint-end slope of the IRLF is much
shallower than the UVLF, deeper observations do not imply that we will
detect higher-redshift samples closer to the detection limit.  In
fact, we find quite the opposite.  The brightest sources at 1.2\,mm,
2\,mm and 3\,mm are expected to sit at much higher redshifts than
their faintest sources.  Finding the highest redshift galaxies
requires wider, shallower surveys.

We examine the measured redshift distributions from 1.2\,mm deep
fields: including the ASPECS-Pilot survey covering 0.79\,arcmin$^2$
\citep{aravena16b}, the HUDF ALMA Deep Field covering 4.4\,arcmin$^2$
\citep{dunlop16a}, and the GOODS-ALMA Deep Field covering
69\,arcmin$^2$ \citep{franco18a}.  The limited statistics of these
surveys (limited to 5, 5, and 15 sources detected above 5$\sigma$,
respectively) do not allow us to draw conclusions as to which of our
three models fits the data best.  The largest dataset from
\citet{franco18a} is significantly limited by their resulting
delivered spatial resolution ($\sim$0.2$''$ beam), which likely led to
severe sample incompleteness for resolved sources \citep[most DSFGs
  are expected to have millimeter sizes $\sim$0.4-0.5$''$ in
  FWHM;][]{hodge16a}.

We also explore measurements of the dust content of UV-selected
galaxies and contrast with the analysis of \citet{bouwens16a} and
\citet{capak15a}.  Both works have claimed that UV-selected
populations appear to be significantly less dusty than expected given
their rest-frame UV colors and/or stellar masses.  We re-assess some
of the base assumptions made in this claim, primarily with the assumed
IR SEDs, and conclude that there is no evidence for less dust in
high-$z$ UV-selected galaxy populations.  Unlike other works in the
literature that claim the UV-selected galaxies require much hotter
dust temperatures to fall in-line with the local IRX--$\beta$
relationships, we find that such hot temperatures are not required to
find consistency from low-$z$ to high-$z$.  We caution that future
conclusions on the dust content of high-$z$ galaxies require a more
thorough analysis of galaxies' IR SEDs, and should ideally not be
limited to a single photometric point.  Similarly, simplistic
assumptions about high-$z$ galaxy SEDs should be replaced with a more
rigorous analysis of SEDs across galaxy populations and environments.

With the goal of pushing our understanding of dusty galaxies toward
the Epoch of Reionization in mind, then a much more optimal strategy
for ALMA would be to map out somewhat wider and shallower 2\,mm
surveys.  The 2\,mm wavelength regime benefits significantly from the
very negative {\it K}-correction \citep*[see][ Figure 3]{casey14a}, in
such a way that filters out lower redshift $z\sim1-3$ sources that
have already been well-characterized in terms of their volume density
and bulk contribution to cosmic star-formation.  We suggest that an
ALMA Band 4 survey of order $\sim$230\,arcmin$^2$ to an RMS of
$\sim$0.08\,mJy and a beamsize $\sim$1--2$''$ will have between
20--120 galaxy detections, a median redshift between $3<z<4.5$, and a
long tail out to very high-$z$.  This dataset would easily distinguish
between competing models for the early Universe IRLF.  Inference of
the IRLF at these epochs will have direct implications for the
prevalence of dusty starbursts during the EoR at $z>6$.  We have also
analyzed the potential of the 3\,mm band to detect dust continuum
sources; though observations require much more depth, as galaxies'
3\,mm flux densities will be intrinsically much lower than at
2\,mm. However, this depth is routinely achieved in observations
intended for molecular line analysis (often the detection of low-J
transitions of CO in moderate redshift galaxies).  An analysis of the
3\,mm number counts already available in the ALMA archive will follow
in Zavala \etal, in preparation.

This paper has shown that the optimum design of ALMA deep fields is
not necessarily obvious, is highly dependent on the driving science
goal, and does not follow the same logic as was used to motivate the
legacy deep field products of the {\it Hubble Space Telescope}, and
soon, the {\it James Webb Space Telescope}.  It is the combination of
the strong negative {\it K}-correction in the submillimeter/millimeter
and the shallow faint-end slope of the IRLF in comparison to the UVLF
(likely caused by the correlation of obscuration with galaxy stellar
mass) that drive the observed differences between {\it HST} deep
fields and ALMA deep fields.  We find that the search for
high-redshift dusty galaxies would be optimized at 2\,mm using
somewhat wider, and shallower survey mapping strategies.  ALMA can
contribute significantly to our constraints on dust in the first few
billion years of the Universe's history through targeted 2\,mm
surveys and large scale analysis and expansion of 3\,mm dual-purpose
archival datasets.

\acknowledgments

The authors wish to thank the Aspen Center for Physics for hosting two
summer workshops, ``The Obscured Universe: Dust and Gas in Distant
Starburst Galaxies'' in summer 2013 and ``New Frontiers in
Far-infrared and Sub-millimeter Astronomy'' in summer 2016, whose
stimulating conversations led to this work.  The Aspen Center for
Physics is supported by National Science Foundation grant PHY-1066293.
CMC thanks the National Science Foundation for support through grant
AST-1714528, and additionally CMC and JAZ thank the University of
Texas at Austin College of Natural Sciences for support.  JS thanks
the McDonald Observatory at the University of Texas at Austin for
support through a Smith Fellowship.  EdC gratefully acknowledges the
Australian Research Council for funding support as the recipient of a
Future Fellowship (FT150100079).  JAH acknowledges support of the VIDI
research programme with project number 639.042.611, which is (partly)
financed by the Netherlands Organisation for Scientific Research
(NWO).  SLF acknowledges support from an NSF AAG award AST-1518183.
This paper makes use of the following ALMA data:
2012.1.00173.S, 2012.1.00756.S, 2013.1.00718.S, 2013.1.00162.S,
2015.1.00543.S, and 2017.1.00755.S.  ALMA is a partnership of ESO
(representing its member states), NSF (USA) and NINS (Japan), together
with NRC (Canada), MOST and ASIAA (Taiwan), and KASI (Republic of
Korea), in cooperation with the Republic of Chile. The Joint ALMA
Observatory is operated by ESO, AUI/NRAO and NAOJ.

Many of the datasets this paper and analyses have only been made
possible by those that were obtained on the summit of Maunakea on the
island of Hawai'i.  The authors wish to recognize and acknowledge the
very significant cultural role and reverence that the summit of
Maunakea has always had within the indigenous Hawaiian community.
Astronomers are most fortunate to have the opportunity to conduct
observations from this mountain.

\bibliography{caitlin-bibdesk}

\end{document}